\documentclass{article}

\usepackage[left=1.5cm,right=1.5cm, top=2.0cm, bottom=2.5cm]{geometry}
\usepackage{graphicx,caption}
\usepackage{relsize}
\usepackage{threeparttable}
\usepackage{ amssymb }
\usepackage{natbib}
\usepackage{multicol,caption}

\newenvironment{Figure1s}
  {\par\medskip\noindent\minipage{\linewidth}}
  {\endminipage\par\medskip}

\usepackage{fancyhdr}
\makeatletter
\let\saved@includegraphics\includegraphics
\AtBeginDocument{\let\includegraphics\saved@includegraphics}
\renewenvironment*{figure}{\@float{figure}}{\end@float}
\makeatother

\def\1E0102{E\,0102}
\def\ci{[C\,\textsc{\smaller I}]}
\def\oiii{[O\,\textsc{\smaller III}]}
\def\oi{O\,\textsc{\smaller I}}
\def\oii{[O\,\textsc{\smaller II}]}
\def\fexiv{[Fe\,\textsc{\smaller XIV}]}
\def\fexi{[Fe\,\textsc{\smaller XI}]}
\def\fex{[Fe\,\textsc{\smaller X}]}
\def\nei{Ne\,\textsc{\smaller I}}
\def\arcsec{$^{\prime\prime}$}

\title{Identification of the Central Compact Object in the young supernova remnant 1E\,0102.2-7219} 

\author
{Fr\'ed\'eric P.A. Vogt$^{1,2,}$\footnote{frederic.vogt@alumni.anu.edu.au} , Elizabeth S. Bartlett$^{1,2}$, Ivo R. Seitenzahl$^{3,4}$, Michael A. Dopita$^{4}$, \\ Parviz Ghavamian$^{5}$, Ashley J. Ruiter$^{3,4,6}$, Jason P. Terry$^{7}$\\
\\
\normalsize{$^{1}$European Southern Observatory, Av. Alonso de C\'ordova 3107, 763 0355 Vitacura, Santiago, Chile}\\
\normalsize{$^{2}$ESO Fellow}\\
\normalsize{$^{3}$School of Physical, Environmental and Mathematical Sciences, University of New South Wales,}\\
\normalsize{ Australian Defence Force Academy, Canberra, ACT 2600, Australia}\\
\normalsize{$^{4}$Research School of Astronomy and Astrophysics, Australian National University, Canberra, ACT 2611, Australia}\\
\normalsize{$^{5}$Department of Physics, Astronomy and Geosciences, Towson University, Towson, MD, 21252, USA}\\
\normalsize{$^{6}$ARC Centre for All-sky Astrophysics (CAASTRO).}\\
\normalsize{$^{7}$Department of Physics and Astronomy, University of Georgia, Athens, GA, USA}\\
}

\date{}

\begin{document} 



%
\maketitle 
\centerline{\textbf{Published in Nature Astronomy}}
\thispagestyle{fancy}


\begin{multicols}{2}

\textbf{
Oxygen-rich young supernova remnants \citep{vandenBergh1988} are valuable objects for probing the outcome of nucleosynthetic processes in massive stars, as well as the physics of supernova explosions. Observed within a few thousand years after the supernova explosion \citep{Winkler1985}, these systems contain fast-moving oxygen-rich and hydrogen-poor filaments visible at optical wavelengths: fragments of the progenitor's interior expelled at a few 1000 km\,s$^{-1}$ during the supernova explosion. Here we report the first identification of the compact object in 1\1E0102.2-7219 in reprocessed Chandra X-ray Observatory data, enabled via the discovery of a ring-shaped structure visible primarily in optical recombination lines of {\nei} and {\oi}. The optical ring, discovered in integral field spectroscopy observations from the Multi Unit Spectroscopic Explorer (MUSE) at the Very Large Telescope, has a radius of $(2.10\pm0.35)$\,\arcsec$\equiv$$(0.63\pm0.11)$\,pc, and is expanding at a velocity of $90.5_{-30}^{+40}$ km\,s$^{-1}$. It surrounds an X-ray point source with an intrinsic X-ray luminosity $L_{i}$ (1.2--2.0 keV)$=(1.4\pm0.2)\times10^{33}$ erg\,s$^{-1}$. The energy distribution of the source indicates that this object is an isolated neutron star: a Central Compact Object akin to those present in the Cas A \citep{Tananbaum1999,Umeda2000,Chakrabarty2001} and Puppis A \citep{Petre1996} supernova remnants, and the first of its kind to be identified outside of our Galaxy.}

\begin{figure*}[htb!]
\centerline{\includegraphics[scale=0.5]{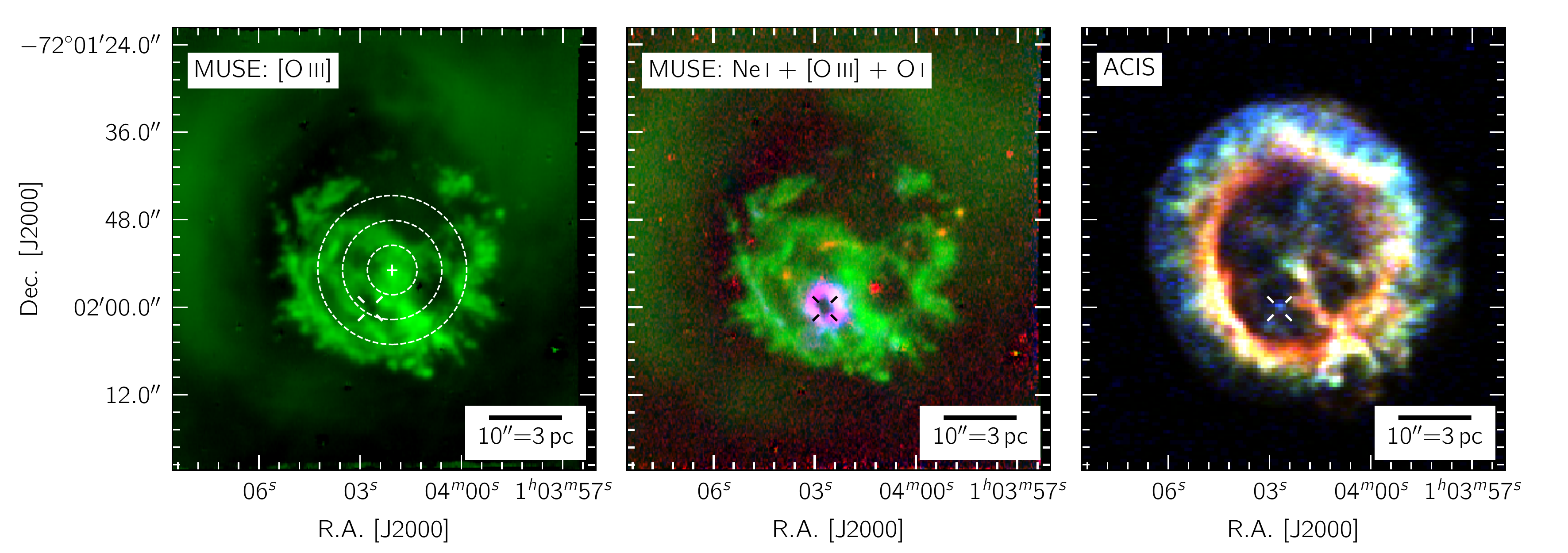}}
\caption{\linespread{1.0}\selectfont{} Global optical and X-ray view of \1E0102. Left: continuum-subtracted MUSE view of \1E0102 in the light of \oiii\,$\lambda\lambda$4959,5007, showing the complex structure of the fast ejecta in this system. The proper motion center of the fast ejecta and its associated 1-$\sigma$, 2-$\sigma$, and 3-$\sigma$ uncertainty areas derived from \textit{HST} observations \citep{Finkelstein2006} is shown using a white "+" and concentric dashed circles. Middle: continuum-subtracted MUSE view of \1E0102 in the light of \nei\,$\lambda$6402 (red channel; 80 km\,s$^{-1}$ velocity range), \oiii\,$\lambda\lambda$4959,5007 (green channel; full velocity range) and \oi\,$\lambda$7774 (blue channel; 80 km\,s$^{-1}$ velocity range). A ring-like structure with a distinct spectral signature is visible in light pink (see Fig.~\ref{fig:zoom} for the zoomed-in, individual \nei\,$\lambda$6402 and \oiii\,$\lambda$5007 images). Right: \textit{CXO} ACIS pseudo-RGB image of \1E0102 (red channel: 0.5-1.2\,keV, green channel: 1.2-2.0\,keV, blue channel: 2.0-7.0\,keV), with a (native ACIS) pixel scale of 0.492\arcsec. A crosshair marker indicates the location of the X-ray point source \textit{p1} in all panels.}\label{fig:wide}
\end{figure*}

SNR 1E\,0102.2-7219 (hereafter referred to as \1E0102) is located in the Small Magellanic Cloud, at a distance of 62\,kpc \citep{Graczyk2014,Scowcroft2016}. It was first identified as an oxygen-rich (O-rich) supernova remnant (SNR) using optical narrow-band imaging \citep{Dopita1981}, on the basis of its X-ray detection by the Einstein Observatory \citep{Seward1981}. The measurement of the O-rich ejecta's proper motions, using \textit{HST} observations spanning an 8-year baseline, indicates an age of 2050$\pm$600\,yr \citep{Finkelstein2006}. The paucity of emission from oxygen-burning products (S, Ca, Ar) originally suggested a Type Ib progenitor \citep{Blair2000}, but the recent detection of [S\,\textsc{\smaller II}]\,$\lambda\lambda$6716,6731 and H$\alpha$ emission in localized, fast knots calls this into question \citep{Seitenzahl2018}. In October 2016, we obtained new observations of \1E0102 with the MUSE optical integral field spectrograph \citep{Bacon2010} mounted on the Nasmyth B of the Unit Telescope 4 of ESO's Very Large Telescope at the observatory of Cerro Paranal in Chile, under Director Discretionary Time program 297.D-5058 (P.I.: Vogt). This dataset (see the Methods for details on the observations and data processing), covers the entire spatial extent of the remnant (see Fig.~\ref{fig:wide}) with a seeing limited resolution of 0.7\arcsec$\equiv$0.21\,pc and a spectral range of 4750--9350\,\AA. The spatio-kinematic complexity of the supernova ejecta is revealed in the MUSE data primarily in the light of \oiii\,$\lambda\lambda$4959,5007. In the coronal lines of \fexiv\,$\lambda$5303, \fexi\,$\lambda$7892 and \fex\,$\lambda$6375, this MUSE dataset also revealed for the first time a thin shell (in emission) surrounding the fast ejecta, tracing the impact of the forward shock wave at optical wavelengths \citep{Vogt2017a}. 

In this Report, we present the discovery of a new structure in \1E0102 revealed by our MUSE observations; a pc-scale low-ionization ``optical ring'' visible (in emission) in 32 recombination lines of \oi\ and \nei\ (see Fig.~\ref{fig:wide}), as well as forbidden lines \ci\,$\lambda$8727, [\oi]\,$\lambda$6300, and [\oi]\,$\lambda$6364. A coincident but spatially more extended high-ionization ring-like structure is visible in the forbidden lines of \oiii\ and \oii. The low-ionization ring --the detailed spectral characterization of which is included in the Supplementary Information-- lacks optical emission from hydrogen and helium, indicating that it is largely composed of heavy elements. Yet, its spectral signature differs from that of the typical fast ejecta in the system. Optical recombination lines such as \nei\,$\lambda$6402, \oi\,$\lambda$7774 and \oi\,$\lambda$8446 dominate in flux over forbidden emission lines, which suggests a low temperature and high density for this structure. This is indicative of different physical conditions \textit{in situ} and/or a different excitation mechanism from that of the O-rich fast ejecta encountering the reverse shock \citep{Sutherland1995}. 

The existence of an X-ray point source in the middle of the low-ionization optical ring yields important clues regarding the exact nature of this structure. We identified this X-ray point source after reprocessing 322.6\,ks of \textit{Chandra X-ray Observatory} (\textit{CXO}) observations of \1E0102 (see Fig.~\ref{fig:wide}). With an X-ray flux of $F$(0.5--7.0\,keV)=$(1.4\pm0.2)\times10^{-14}$ erg\,cm$^{-2}$\,s$^{-1}$ (see the Methods for details), the X-ray point source is located at the position:
\begin{equation}
\textrm{R.A: 01$^h$04$^m$02.7$^s$ | Dec: -72$^{\circ}$02$^{\prime}$00.2$^{\prime\prime}$ [J2000].}\nonumber
\end{equation}

The estimated absolute uncertainty of this position is 1.2\arcsec, stemming both from the accuracy of the World Coordinate System (WCS) solution of the combined \textit{CXO} observations, and the complex X-ray background of \1E0102. One can expect $\sim$70 background X-ray sources (e.g. AGNs) with $F$(0.5--2.0\,keV)$>1.0\times10^{-14}$ erg\,cm$^{-2}$\,s$^{-1}$ per degree square \citep{Gilli2007}. Given that the area of the elliptical structure seen by MUSE is $\sim$15 square arcseconds, the probability of this X-ray source to lie in the background (i.e. a \textit{chance alignment}) is extremely low, with only $8\times10^{-5}$ similar (or brighter) sources expected for this area. The spatial coincidence of the optical ring and the X-ray point source is thus a strong indication that the X-ray point source is located in the SMC, and directly associated with the elliptical structure itself.

An earlier, dedicated search for a compact object in \1E0102 \citep{Rutkowski2010} already located the same X-ray point source described above. For consistency, we shall thus refer to this X-ray source with the same name used by the earlier search: ``\textit{p1}''\footnote{Although \cite{Rutkowski2010} never explicitly quote the coordinates of \textit{p1}, its location is made clear from their Fig.~1}. This source was then merely listed as one candidate among seven sources within \1E0102, but with no explicit comment. We hypothesize that source \textit{p1} may have (then) escaped a definitive identification because of its embedment in the diffuse and confusing X-ray background in the interior of \1E0102, and/or because it was located outside of the ``test source region''\footnote{The ``test source region" was defined as the angular area where a compact object displaced by a natal kick would be most likely to be found. However, this area was computed using only a mean pulsar kick velocity. It was also less conservative than it could have been if the measurement errors associated with the explosion center had been included.} defined by \cite{Rutkowski2010}. Today, it is the unique capabilities of MUSE (i.e. its high sensitivity, fine spatial sampling and R$\cong$3000 spectral resolution) that make it possible to spatio-kinematically isolate a distinctive optical ring of ejecta material centered on the X-ray source, thereby allowing us to firmly associate \textit{p1} with the SNR.  

Evidently, the existence of an X-ray point source physically associated with \1E0102 raises the question: could \textit{p1} be the as-of-yet unidentified compact object leftover by the supernova? In the other O-rich SNRs, compact objects come in two distinct types: (a) pulsars with active pulsar wind nebulae, as detected in G292.0+1.8 \citep{Camilo2002,Park2007} and 0540-69.3 \citep{Seward1984,Middleditch1985,Mignani2010}, and (b) Central Compact Objects (CCOs), as detected in Puppis A \citep{Petre1996} and Cas A \citep{Chakrabarty2001,Mereghetti2002}. CCOs lack evidence of a pulsar wind nebula. They are detected only via their blackbody radiation, and are understood to be isolated, cooling neutron stars with thermal X-ray luminosities $L\cong10^{33.5}$ erg\,s$^{-1}$ \citep{Vigano2013}. Extensive optical searches have failed to find any counterparts to the CCOs in Cas A and Puppis A, setting strong constraints on the possible accretion rate of material from the immediate surroundings of these objects \citep{Wang2007,Mignani2009}.

We searched for an optical counterpart to \textit{p1} in archival images from the Wide-Field Camera 3 (WFC3) and Advanced Camera for Survey (ACS) on the \textit{Hubble Space Telescope} (\textit{HST}). We found no suitable candidate down to an apparent magnitude upper limit of $m_\textrm{F775W}>24.5$ (see the Methods for details). For completeness, what is then the probability that the source \textit{p1} is a chance encounter of an X-ray source \textit{within the SMC itself}? The SMC is host to a large number of High-Mass X-ray Binaries (HMXB), due in part to its star forming history. After a background AGN, this is statistically the most likely X-ray source one could encounter in the SMC \citep{Sturm2013}. However, the archival \textit{HST} observations of the area rule out the presence of a high mass stellar counterpart \citep{Haberl2016} and thus the HMXB scenario. The lack of an optical counterpart also rules out a chance alignment with an X-ray-bright foreground star. The source \textit{p1} is visible above $\sim$1~keV, so that it is evidently not a Super-Soft X-ray Source: a class of X-ray sources that includes cataclysmic variables, planetary nebulae and accreting white dwarfs. The source \textit{p1} is also incompatible with isolated white dwarfs, which have soft X-ray spectra well fit by blackbody models with $kT$ on the order of eV rather than keV (as is the case of \textit{p1}, to be demonstrated shortly). The last remaining candidate for a chance alignment is that of a quiescent low-mass X-ray binary (LMXBs). To date, there is no known LMXB in the SMC, the number of which scales as a function of mass in a given galaxy \citep{Gilfanov2004}. We expect a maximum of $\sim$35 LMXBs in the SMC, leading to an LMXB density --and thus the probability of a chance alignment-- of $\sim3\times10^{-7}$ over the area of the optical ring. We are thus left with the only plausible scenario: that of \textit{p1} being the compact object of \1E0102.

The low counts associated with the \textit{CXO} observations of the source \textit{p1} hinder our ability to perform a detailed direct spectral analysis. Instead, we perform a ``goodness of fit'' analysis for a series of a) absorbed, single blackbody models and b) absorbed power-laws (see the Methods for details). In the first case, we find that the source \textit{p1} can be best represented by single blackbody model with $kT_\textrm{BB}=(0.19\pm0.02)$\,keV. In the second case, we find that \textit{p1} is best represented by a power law with a photon index $3.75< \Gamma <4.75$. Pulsars and pulsar wind nebulae can typically be described by power-laws with photon index $1\lesssim \Gamma \lesssim2$ \citep{Kargaltsev2008}: a range incompatible with the X-ray spectral energy distribution of the source \textit{p1}, and thus indicating that \textit{p1} is most certainly not a pulsar. From our best-match single black-body model, we derive an intrinsic (unabsorbed) luminosity for the source \textit{p1} of $L_{i}$ (1.2--2.0 keV)$=(1.4\pm0.2)\times10^{33}$ erg\,s$^{-1}$ and a (blackbody equivalent) source radius of 8.7$^{+4.9}_{-2.7}$\,km. Altogether, these observational characteristics (X-ray brightness and temperature, X-ray energy distribution consistent with a soft thermal-like spectrum, lack of an optical counterpart and pulsar wind nebula, and spatial location within a SNR) lead us to propose the X-ray source \textit{p1} as a new addition to the CCO family \citep{Pavlov2004}; the first identified extragalactic CCO.

For comparison purposes, we simulate how the CCO of Cas A, whose X-ray signature has been extensively modelled \citep{Pavlov2009}, would appear if it were located at the distance of the SMC (see the Methods for details). The corresponding spectrum is shown in the top panel of Fig.~\ref{fig:xray}. Our best match single-blackbody model is shown in the bottom panel of the same Figure. The Cas A-like CCO spectrum is 0.2\,keV hotter than that of the source \textit{p1}, but the brightness of both remain comparable. We note that similarly to \textit{p1}, fitting the Cas A CCO with a power law also leads to a large photon index \citep{Pavlov2009}.

\begin{figure*}[htb!]
\vspace{1pt}
\centerline{\includegraphics[scale=0.5]{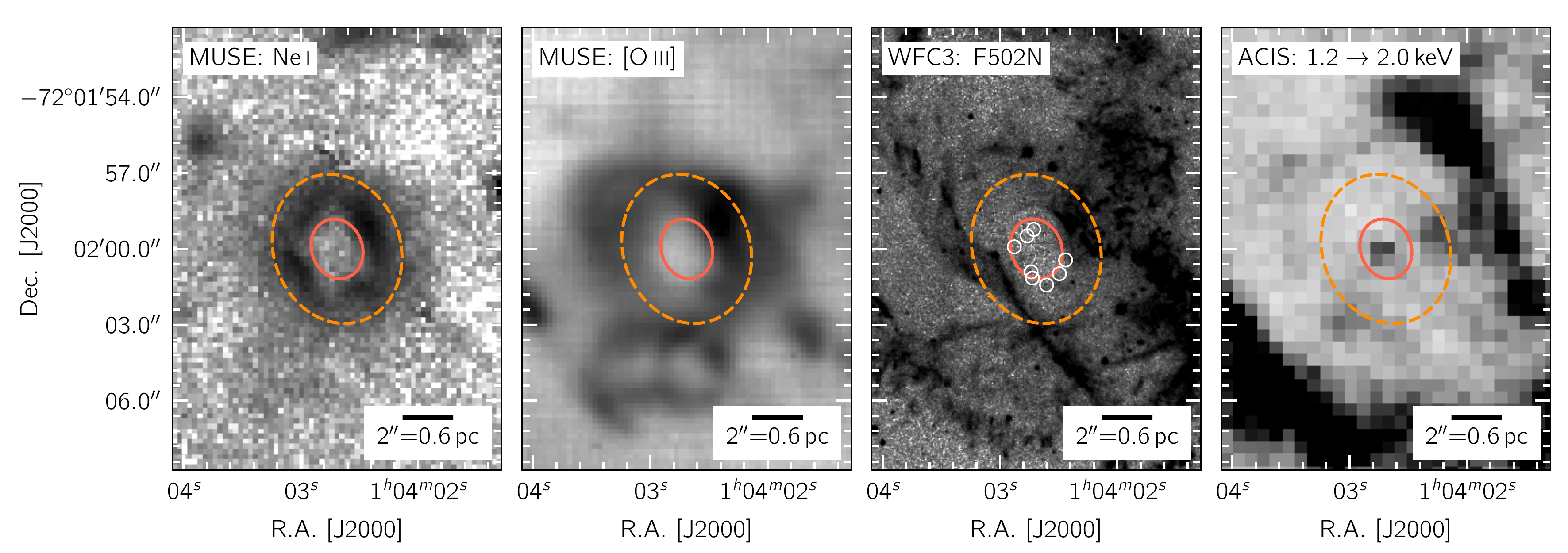}}
\caption{\linespread{1.0}\selectfont{}Close-up view of the optical ring and CCO. Left: Close-up, continuum-subtracted MUSE view of the optical low-ionization ring in the light of \nei\,$\lambda$6402 (80 km\,s$^{-1}$ bandpass). Middle-left: idem, in the light of \oiii\,$\lambda\lambda$5007 (80 km\,s$^{-1}$ bandpass). A large ring-like structure is detected, despite the presence of some contaminating O-rich knots visible via their (redshifted) \oiii\,$\lambda$4959 emission. Middle-right: close-up \textit{HST} ACS F475W view of the same area. The complex structures of the fast ejecta are readily visible. So is a smooth elliptical arc spatially coherent with the ring identified with MUSE. All confirmed optical point-sources in the vicinity of the torus center are marked with white circles. Right: \textit{CXO} ACIS view of the same area in the 1.2--2.0\,keV band, revealing the X-ray point source spatially coincident with the center of the optical gas ring revealed by MUSE. Two ellipses inclined at 20$^{\circ}$ East-of-North with an axis ratio of 1.2, semi-major axis of 1.2\arcsec\ and 3.0\arcsec, and centered at R.A.: 01$^{h}$04$^{m}$02.7$^{s}$; Dec.: -72$^{\circ}$02$^{\prime}$00.2$^{\prime\prime}$ trace the inner and outer edge of the torus in all panels.\newline }\label{fig:zoom}
\end{figure*}

\begin{figure*}[htb!]
\vspace{0pt}
\centerline{\includegraphics[scale=0.5]{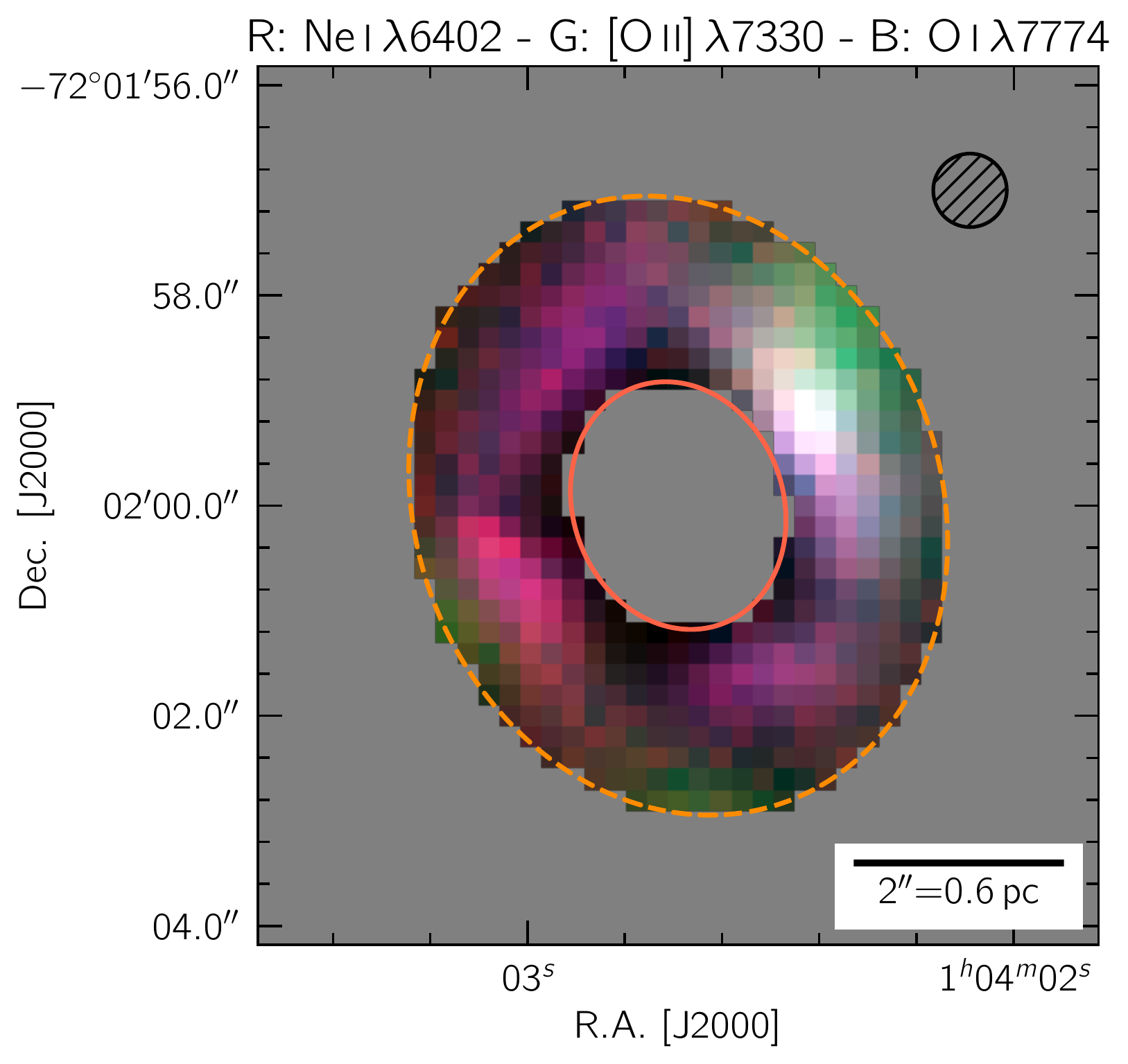}\qquad\includegraphics[scale=0.5]{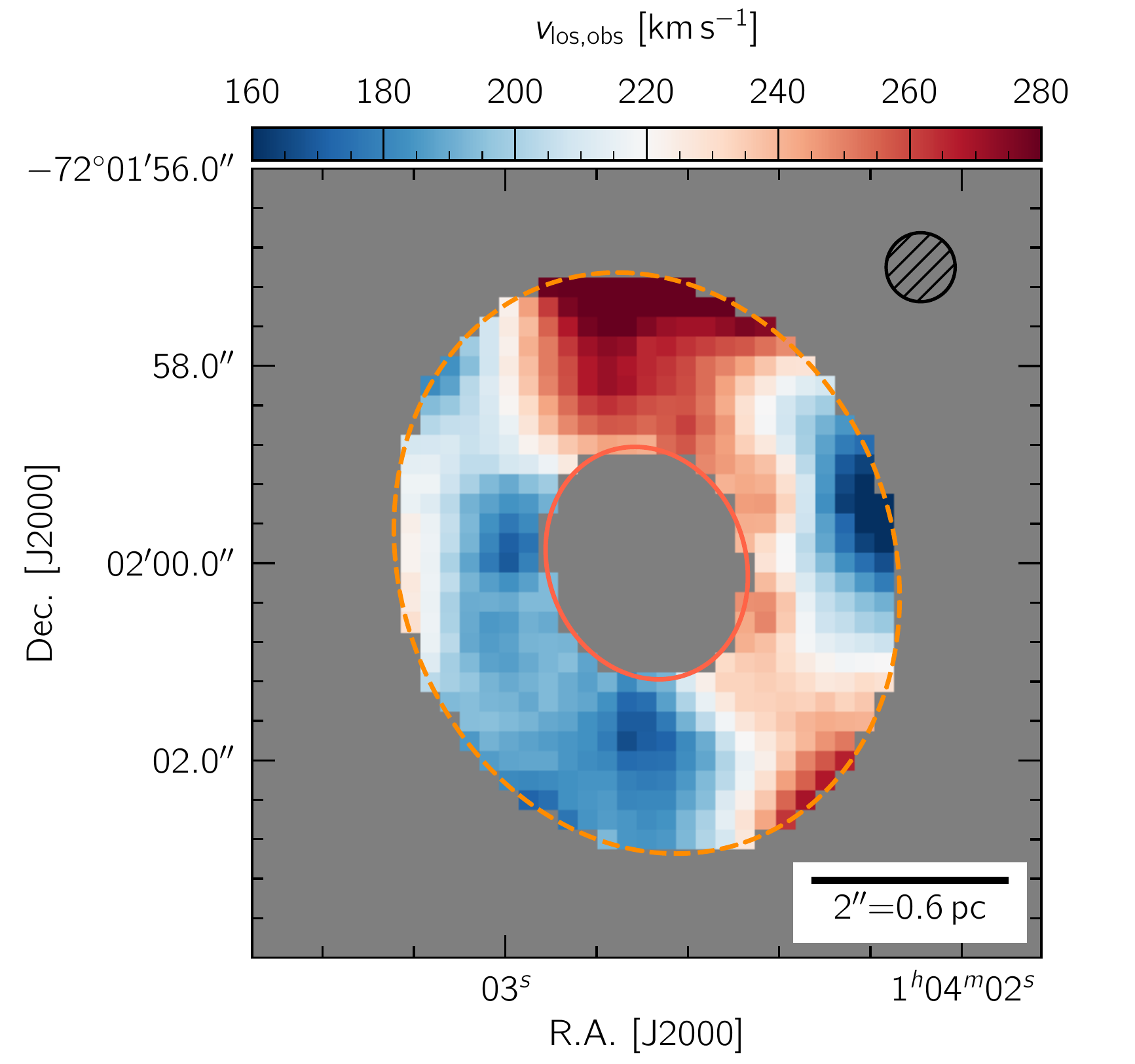}}
\caption{\linespread{1.0}\selectfont{} Spectral analysis of the optical ring. Left: pseudo-RGB image of the optical ring in the light of \nei\,$\lambda$6402 (red channel), \oii\,$\lambda$7330 (green channel) and \oi\,$\lambda$7774 (blue channel). The low-ionization, denser material appears in pink. Lower-density, higher-ionization material (bright in \oii\,$\lambda$7330) is visible (in green) immediately outside of the low-ionization emission towards the NW, SE and S (see also the \oiii\,$\lambda$5007 emission in Fig.~\ref{fig:zoom}). This image is reconstructed using the fitted line fluxes for each spaxel. The spatial resolution of the data (set by the seeing during the MUSE observations) is indicated by the dashed circle. The full and dashed ellipses trace the spatial extent of the optical ring, as indicated in Fig.~\ref{fig:zoom}. Right: observed LOS velocity of the gas, derived for each spaxel from their full spectral fit. The \textit{blueshifted trough} to the West is present in a region of low surface brightness for the low-ionization lines, and the kinematic fit might be influenced by contaminating emission from fast ejecta in this area. Overall, the kinematic pattern is suggestive of an expanding torus, tilted with respect to the plane-of-the-sky. Small-scale variations should not be over-interpreted, given the spatial resolution of this dataset, and the typical MUSE kinematic fitting artefacts, visible to the trained eye as vertical and horizontal jumps \citep{Weilbacher2015,Vogt-thesis}. }\label{fig:torus_RGB}
\vspace{17pt}
\end{figure*}

We now focus on the nature of the optical ring surrounding the newly identified CCO in \1E0102. The elliptical structure has a semi-major axis of $(2.10\pm0.35)$\,\arcsec$\equiv$$(0.63\pm0.11)$\,pc, an ellipticity $b/a=(1.20\pm0.05)^{-1}$, and a ring-width of $(1.80\pm0.35)$\,\arcsec$\equiv(0.54\pm0.11)$\,pc (see Fig.~\ref{fig:zoom}). We find evidence of (at least) four intensity discontinuities along the ring to the N, S, E and S-W (see Fig.~\ref{fig:torus_RGB}). Unambiguously identifying sub-structures within the ring itself will require sharper follow-up observations. In the \textit{HST} WFC3 F502N image presented in Fig.~\ref{fig:zoom}, we note that a thin elliptical arc is present near the inner edge of the area coincident with the optical ring seen by MUSE, with a semi-major axis of $(1.65\pm0.04)$\,\arcsec$\equiv$$(0.50\pm0.01)$\,pc. Given the spectral transmission window of the F502N filter, this arc is most certainly detected via its \oiii\,$\lambda\lambda$4959,5007 emission. However, unambiguously separating it from the complex structure of the O-rich fast ejecta in this area is not straightforward. Our MUSE observations suggest that the high-ionization ring seen in \oiii\,$\lambda\lambda$4959,5007 may be connected to a larger, funnel-like structure of shocked ejecta oriented along the line of sight.

\begin{Figure1s}
\centerline{\includegraphics[scale=0.6]{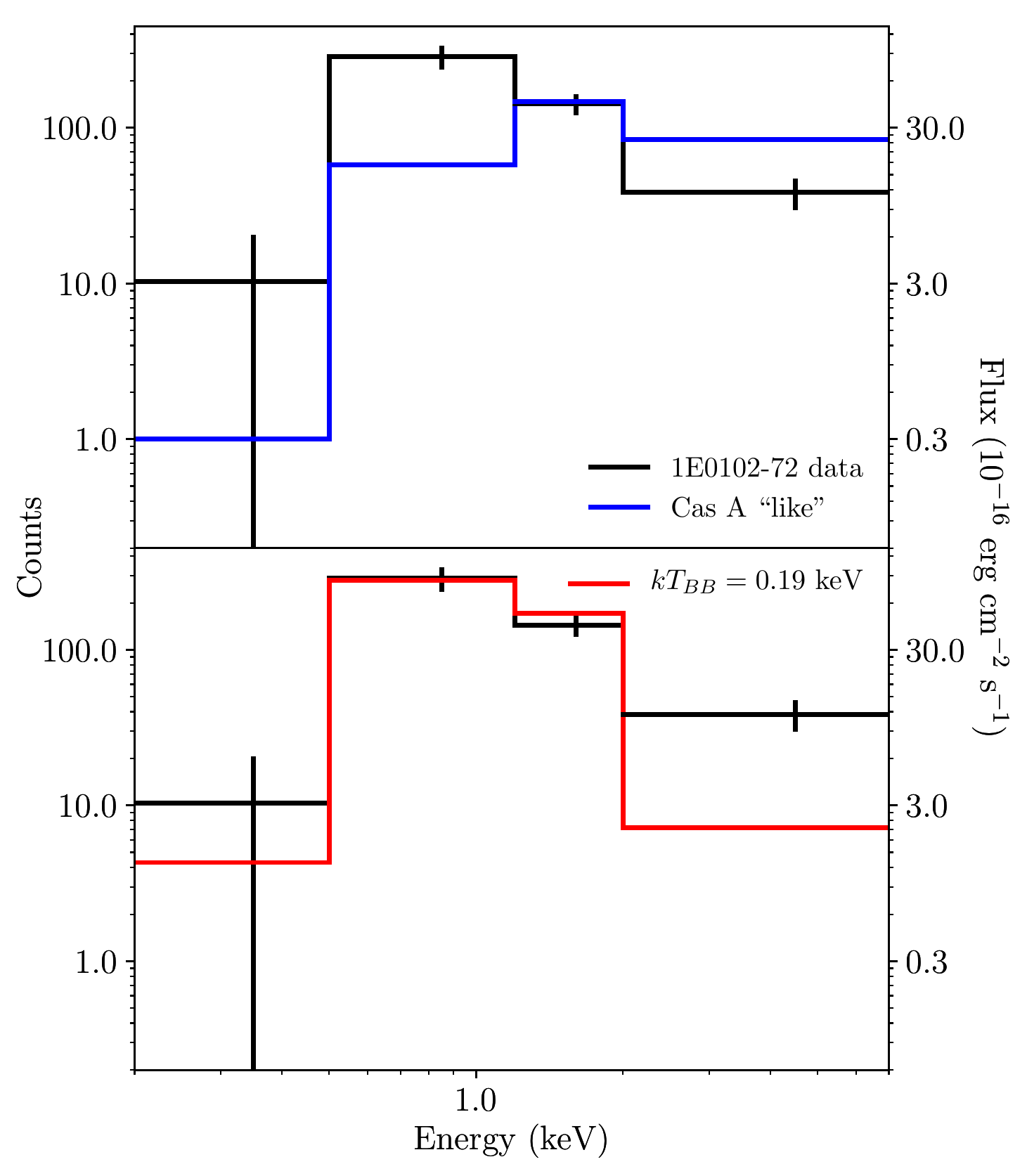}}
\captionof{figure}{\linespread{1.0}\selectfont{}X-ray spectral signature of \textit{p1}, and associated modelling. Energy distribution of the X-ray point source \textit{p1} in \1E0102 (black curve), compared with that of a Cas A-like CCO observed at the same distance (blue curve, top), and our best-match, absorbed, single blackbody component model with $kT_\textrm{BB}=0.19$\,keV (red curve, bottom). Given its brightness and spectral signature, \textit{p1} is consistent with being a CCO, $\sim$0.2\,keV cooler than that of Cas A \citep{Pavlov2009}.}\label{fig:xray}
\end{Figure1s} 

We performed a full spectral fit of the low-ionization optical ring (on a spaxel-by-spaxel basis, see the Methods for details), and derived its line-of-sight (LOS) kinematics (presented in Fig.~\ref{fig:torus_RGB}). We find a clear asymmetry of $\Delta v_\textrm{los}=(100\pm21)$ km\,s$^{-1}$ for the gas LOS velocity, broadly aligned with the ring's major axis rotated by $(20\pm5)^{\circ}$ East-of-North. The radius of the low-ionization ring defined relative to the CCO implies an escape velocity of $\lesssim1$ km\,s$^{-1}$: given its LOS velocity, the ring material is thus not gravitationally bound to the CCO.  Despite an evident \textit{blueshifted trough} to the West (possibly due to intervening fast ejecta in the line of sight biasing the measurements of the line kinematics), the overall pattern is suggestive of an expanding torus, tilted with respect to the plane-of-the-sky, by $33^{\circ}.5_{-4.5}^{+3.5}$, with an absolute expansion velocity of $90.5_{-30}^{+40}$ km\,s$^{-1}$. If the torus has been expanding ballistically, this would imply an age of $6800^{+5400}_{-2900}$\,yr for the structure. This age is somewhat older than (though not incompatible with) the supernova age derived from the proper motion of the O-bright fast ejecta \citep{Finkelstein2006}, given the associated uncertainties. We discourage any over-interpretation of the arcsec-scale structures of the ring velocity map, recalling both the limited spatial resolution of the observations, and the existence of typical kinematic fitting artefacts associated with MUSE datacubes \citep{Weilbacher2015,Vogt-thesis}.  

The exact physical mechanism(s) responsible for the excitation of the low-ionization ring remains uncertain. A spectrum dominated by recombination lines of {\nei} and {\oi} is not a proof that the ring is solely composed of Ne and O, but it does indicate that its electron temperature is $\lesssim$3000\,K. Altogether, the optical spectrum and kinematics of the ring are consistent with dense, photoionized material which has not yet passed through the reverse shock.  Quantifying the respective influence of photoionization by the CCO and/or by the overlying reverse shocked ejecta \citep[for example, as modelled for SN\,1006;][]{Hamilton1988} will require theoretical modelling beyond the scope of this Letter.  

The existence of a slowly expanding torus of low-ionization material surrounding the CCO of {\1E0102} leads us to propose this location as the actual supernova explosion site of the system. If ejected during the supernova explosion, the slowly expanding material in the torus (in sharp contrast with the large velocities measured elsewhere in \1E0102) would have originated from close to the supernova mass cut: the surface separating ejected material from material that forms the CCO \citep{Umeda2016, Hix2016}. But we also note that our current kinematic measurements do not rule out the possibility that this structure could pre-date the supernova explosion by a few kyr. Either way, given that the CCO is still located within $\sim$1.2\arcsec\ from center of the optical ring 2050\,yr after the supernova explosion, we can set an upper-limit on the transverse velocity of the CCO of $\lesssim$170 km\,s$^{-1}$. In this scenario, our newly defined explosion center would be located 5.9\arcsec($\equiv$1.77\,pc) from the explosion center derived from the fast ejecta proper motion \citep[Fig.\ref{fig:wide} and ][]{Finkelstein2006} -- well within the 2-$\sigma($=6.8\arcsec) uncertainty associated with that measurement. Setting the explosion site of \1E0102 at the current location of the CCO would however imply a large offset with respect to the rather regular outer envelope of the X-ray emission of this SNR, requiring a specific set of circumstances to shape the expansion of the forward shock wave into such a regular structure. For example, the forward shock may be expanding inside a cavity carved by pre-supernova winds, which combined with a density gradient in the surrounding interstellar medium could induce an offset between the center of the X-ray shell and the explosion site.

$ $\\
$ $\\
In the alternative scenario, the supernova explosion occurred elsewhere, away from the current location of the CCO. Under these circumstances, we may be observing a CCO with high transverse velocity -e.g. $\sim$850\,km\,s$^{-1}$ in the plane-of-the-sky assuming the explosion center from \cite{Finkelstein2006}- as it catches up and collides (possibly) with surrounding ejecta. But given the timescales involved, we fail, at this point in time, to identify a physical mechanism able to account for the dimensions and kinematics of the low-ionization ring under these circumstances. A re-analysis of existing archival \textit{HST} observations of \1E0102 spanning $>$8\,yr should help refine the measurement of the explosion center from the kinematics of the fast ejecta in \1E0102. If precise enough, such a refined measurement can provide a strong test of our presently favored scenario: that of the supernova explosion originating at the current location of the CCO. At the same time, follow-up observations of the ring surrounding the CCO, with higher spatial resolution, are warranted to resolve the sub-structures within the ring, link the low-ionization material to the elliptical \oiii\ arc detected in \textit{HST} images of the area, and refine the derivation of its kinematic signature and age. The Adaptive Optics modes of MUSE appear particularly suited to the task.  

\paragraph*{Online Content} Methods, along with additional Supplementary Information display items, are available in the online version of the paper; references unique to these sections appear only in the online paper.

{\smaller
\setlength{\bibsep}{1pt plus 0.3ex}
\bibliography{bibliography_fixed}
\bibliographystyle{aa}
}

\section*{Acknowledgments}
We thank Eric M. Schlegel and the other two (anonymous) reviewers for their constructive comments. This research has made use of \textsc{brutus}, a Python module to process data cubes from integral field spectrographs hosted at \newline \textit{http://fpavogt.github.io/brutus/}. For this analysis, \textsc{brutus} relied on \textsc{statsmodel} \citep{Seabold2010}, \textsc{matplotlib} \citep{Hunter2007}, \textsc{astropy}, a community-developed core Python package for Astronomy
\citep{AstropyCollaboration2013}, \textsc{aplpy}, an open-source plotting package for Python \citep{Robitaille2012}, and \textsc{montage}, funded by the National Science Foundation under Grant Number ACI-1440620 and previously funded by the National Aeronautics and Space Administration's Earth Science Technology Office, Computation Technologies Project, under Cooperative Agreement Number NCC5-626 between NASA and the California Institute of Technology. 

This research has also made use of \textsc{drizzlepac}, a product of the Space Telescope Science Institute, which is operated by AURA for NASA, of the \textsc{aladin} interactive sky atlas \citep{Bonnarel2000}, of \textsc{saoimage ds9} \citep{Joye2003} developed by Smithsonian Astrophysical Observatory, of NASA's Astrophysics Data System, and of the NASA/IPAC Extragalactic Database \citep{Helou1991} which is operated by the Jet Propulsion Laboratory, California Institute of Technology, under contract with the National Aeronautics and Space Administration. 

This work has made use of data from the European Space Agency (ESA) mission {\it Gaia} \newline (\textit{https://www.cosmos.esa.int/gaia}), processed by the {\it Gaia} Data Processing and Analysis Consortium (DPAC, \newline \textit{https://www.cosmos.esa.int/web/gaia/dpac/consortium}). Funding for the DPAC has been provided by national institutions, in particular the institutions participating in the {\it Gaia} Multilateral Agreement. Some of the data presented in this paper were obtained from the Mikulski Archive for Space Telescopes (MAST). STScI is operated by the Association of Universities for Research in Astronomy, Inc., under NASA contract NAS5-26555. Support for MAST for non-HST data is provided by the NASA Office of Space Science via grant NNX09AF08G and by other grants and contracts.

IRS was supported by Australian Research Council Grant FT160100028. PG acknowledges support from \textit{HST} grant HST-GO-14359.011. AJR has been funded by the Australian Research Council grant numbers CE110001020 (CAASTRO) and FT170100243. FPAV and IRS thank the CAASTRO AI travel grant for generous support. PG thanks the Stromlo Distinguished Visitor Program.  

Based on observations made with ESO Telescopes at the La Silla Paranal Observatory under program ID 297.D-5058.

\section*{Author Contributions}
F.P.A.V. reduced and lead the analysis of the MUSE datacube. E.S.B. lead the spectral analysis of the Chandra dataset. All authors contributed to the interpretation of the observations, and the writing of the manuscript.

\section*{Author Information} 
The authors declare no competing financial interests. Readers are welcome to comment on the online version of the paper. Correspondence and requests for materials should be addressed to F.P.A.V. (frederic.vogt@alumni.anu.edu.au).

\clearpage

\section*{Methods}

\subsection*{Observations, data reduction \&\\ post-processing}\label{sec:obs}
\subsubsection*{MUSE}
The MUSE observations of \1E0102 acquired under Director Discretionary Time program 297.D-5058 (P.I.: Vogt) are comprised of nine 900\,s exposures on-source. The detailed observing strategy and data reduction procedures are described exhaustively in \cite{Vogt2017a}, to which we refer the interested reader for details. Similarly to that work, the combined MUSE cube discussed in the present report was continuum-subtracted using the Locally Weighted Scatterplot Smoothing algorithm \citep{Cleveland1979}. This non-parametric approach is particularly suitable to reliably remove both the stellar and nebular continuum in all spaxels of the datacube without the need for manual interaction. 

The one major difference between the MUSE datacube of \1E0102 described in \cite{Vogt2017a} and this work lies in the World Coordinate System (WCS) solution. For this analysis, we have refined the WCS solution (then derived by comparing the MUSE white-light image with the Digitized Sky Survey 2 red image of the area) by anchoring it to the \textit{Gaia} \citep{GaiaCollaboration2016a} Data Release 1 \citep[DR1][]{GaiaCollaboration2016} entries of the area. In doing so, we estimate our absolute WCS accuracy to be of the order of 0.2\arcsec.

\subsubsection*{\textit{CXO}}
\1E0102 has been used as an X-ray calibration source for many years \citep{Plucinsky2017}, so that there exist numerous \textit{CXO} observations of this system in the archive. When selecting datasets to assemble a deep X-ray view of \1E0102, we applied the following, minimal selection criteria:
\begin{itemize}
\item \textsc{datamode = vfaint}
\item no CC-mode observations
\item $\left|\textsc{fp\_temp}-153.3 \right|<2^{\circ}$
\item \textsc{sepn} $\leq$ 1.2\,arcmin, with the reference point set to R.A.:01$^{h}$04$^{m}$02$^{s}$.4, Dec.:-72$^{\circ}$01$^{\prime}$55$^{\prime\prime}$.3 [J2000].
\end{itemize}
These criteria ensure that we combine a uniform set of observations with minimal instrumental background. The last condition ensures that we use the observations with the highest possible spatial resolution, by using only those pointings with \1E0102 close from the optical axis of the telescope. The resulting list of 28 observations matching these criteria is presented in Table~\ref{table:cxo}. We did not apply any selection criteria on the year or the depth of the observations: a direct consequence of setting the focus of our analysis on the characterization of the detected X-ray point source (for which we find no evidence of proper motion in the \textit{CXO} observations), rather than on the fast-moving ejecta. 

We fetched and reprocessed all these datasets using \textsc{ciao 4.9} and \textsc{caldb 4.7.3}, via the \textsc{chandra\_repro} routine. We searched for evidence of background flares by generating images in the 0.5-8.0\,keV energy range while removing the bright sources in the field of view, including \1E0102. Light curves were then created from the entire remaining image and inspected by eye. We dropped three observations (Obs. I.D.: 5123, 5124 and 6074) that show signs of flaring, prior to combining the remaining ones using the \textsc{reproject\_obs} routine. All X-ray images of \1E0102 presented in this work were extracted from this combined, reprojected dataset. We did not perform any adjustment to the existing WCS solution of the combined dataset, which given our \textit{target on-axis} selection criteria can be expected to be of the order of 0.6\arcsec. 

\begin{table*}
\begin{center}
{
\caption{ List of \textit{CXO} observations of \1E0102 reprocessed for this work. All these datasets but three (subject to flaring) were combined and analyzed jointly.}\label{table:cxo}\smaller
\begin{tabular}{c c c c}
\hline
Obs. I.D. & Date & Depth & Signs of flaring ?\\
      & & [ks] & \\
\hline
\hline
3519  & 2003-02-01  &  8.0 &  \\
3520  & 2003-02-01  &  7.6 &  \\
3545  & 2003-08-08  &  7.9 &  \\
3544  & 2003-08-10  &  7.9 &  \\
5123  & 2003-12-15  & 20.3 & yes  \\
5124  & 2003-12-15  &  7.9 & yes  \\
5131  & 2004-04-05  &  8.0 &  \\
5130  & 2004-04-09  & 19.4 &  \\
6075  & 2004-12-18  &  7.9 &  \\
6042  & 2005-04-12  & 18.9 &  \\
6043  & 2005-04-18  &  7.9 &  \\
6074  & 2004-12-16  & 19.8 & yes \\
6758  & 2006-03-19  &  8.1 &  \\
6765  & 2006-03-19  &  7.6 &  \\
6759  & 2006-03-21  & 17.9 &  \\
6766  & 2006-06-06  & 19.7 &  \\
8365  & 2007-02-11  & 21.0 &  \\
9694  & 2008-02-07  & 19.2 &  \\
10654 & 2009-03-01  &  7.3 &  \\
10655 & 2009-03-01  &  6.8 &  \\
10656 & 2009-03-06  &  7.8 &  \\
11957 & 2009-12-30  & 18.4 &  \\
13093 & 2011-02-01  & 19.0 &  \\
14258 & 2012-01-12  & 19.0 &  \\
15467 & 2013-01-28  & 19.1 &  \\
16589 & 2014-03-27  &  9.6 &  \\
18418 & 2016-03-15  & 14.3 &  \\
19850 & 2017-03-19  & 14.3 &  \\
\hline
\end{tabular}}
\end{center}
\end{table*}

\subsubsection*{\textit{HST}}

We downloaded from the Barbara A. Mikulski Archive for the Space Telescopes (MAST) all the observations of \1E0102 acquired with the Advanced Camera for Survey (ACS) and the Wide Field Camera 3 (WFC3) that used narrow, medium and wide filters. These belong to three observing programs: 12001 (ACS, P.I.: Green), 12858 (ACS, P.I.: Madore) and 13378 (WFC3, P.I.: Milisavljevic). The exhaustive list of all the individual observations is presented in Table~\ref{table:hst}.

\begin{table*}
\begin{center}
{
\caption{List of \textit{HST} observations of \1E0102 reprocessed for this work.}\label{table:hst}
\smaller
\begin{tabular}{c c c c c c c}
\hline
Program I.D. & Observation I.D. & P.I. & Observation Date & Instrument & Filter & Exposure time\\
      & & & & & & [s] \\
\hline
\hline

13378 & ICBQ01010 & Milisavljevic & 2014-05-12 & WFC3/UVIS & F280N  & 1650 \\
13378 & ICBQ01020 & Milisavljevic & 2014-05-12 & WFC3/UVIS & F280N  & 1100 \\
13378 & ICBQ01050 & Milisavljevic & 2014-05-12 & WFC3/UVIS & F373N  & 2268 \\
13378 & ICBQ03070 & Milisavljevic & 2014-05-14 & WFC3/UVIS & F467M  & 1362 \\
12001 & J8R802010 & Green         & 2003-10-15 & ACS/WFC   & F475W  & 1520 \\
12001 & J8R802011 & Green         & 2003-10-15 & ACS/WFC   & F475W  & 760 \\
12858 & JBXR02010 & Madore        & 2013-04-10 & ACS/WFC   & F475W  & 2044 \\
13378 & ICBQ02010 & Milisavljevic & 2014-05-13 & WFC3/UVIS & F502N  & 1653 \\ 
13378 & ICBQ02020 & Milisavljevic & 2014-05-13 & WFC3/UVIS & F502N  & 1100 \\
12001 & J8R802020 & Green         & 2003-10-15 & ACS/WFC   & F550M  & 1800 \\
12001 & J8R802021 & Green         & 2003-10-15 & ACS/WFC   & F550M  & 900 \\
13378 & ICBQ03080 & Milisavljevic & 2014-05-14 & WFC3/UVIS & F645N  & 1350 \\
13378 & ICBQ03010 & Milisavljevic & 2014-05-14 & WFC3/UVIS & F657N  & 1653 \\
13378 & ICBQ03020 & Milisavljevic & 2014-05-14 & WFC3/UVIS & F657N  & 1102 \\
12001 & J8R802030 & Green         & 2003-10-15 & ACS/WFC   & F658N  & 1440 \\
12001 & J8R802031 & Green         & 2003-10-15 & ACS/WFC   & F658N  & 720 \\
13378 & ICBQ03030 & Milisavljevic & 2014-05-14 & WFC3/UVIS & F665N  & 1719 \\
13378 & ICBQ03040 & Milisavljevic & 2014-05-14 & WFC3/UVIS & F665N  & 1146 \\
13378 & ICBQ03050 & Milisavljevic & 2014-05-14 & WFC3/UVIS & F673N  & 1719 \\
13378 & ICBQ03060 & Milisavljevic & 2014-05-14 & WFC3/UVIS & F673N  & 1146 \\
12001 & J8R802050 & Green         & 2003-10-15 & ACS/WFC   & F775W  & 1440 \\
12001 & J8R802051 & Green         & 2003-10-15 & ACS/WFC   & F775W  & 720 \\
12001 & J8R802040 & Green         & 2003-10-15 & ACS/WFC   & F850LP & 1440 \\
12001 & J8R802041 & Green         & 2003-10-15 & ACS/WFC   & F850LP & 720 \\
\hline
\end{tabular}}
\end{center}
\end{table*}

All the calibrated, CTE-corrected, individual exposures (\textsc{*\_flc.fits}) obtained from MAST were fed to the \textsc{tweakreg} routine to correct their WCS solutions. We used a custom \textsc{python} script relying on the \textsc{drizzlepac 2.1.13}, \textsc{astroquery} and \textsc{astropy} packages to do so automatically for all the filters. We used the \textit{Gaia} \citep{GaiaCollaboration2016a} DR1\citep{GaiaCollaboration2016} catalogue as the reference set of point source coordinates and fluxes. Using the same script, we subsequently fed all the images for a given filter to the \textsc{astrodrizzle} routine, and combined them into a single final frame. We set the pixel scale to 0.04\arcsec\ for all filters for simplicity, but note that this value does not affect significantly our conclusions. We also note that combining observations separated by several years is not ideal from the perspective of the fast ejecta (whose proper motion imply a blurring of the final image). It is however a suitable approach in the present case, i.e. in order to look for an optical counterpart to the X-ray point source detected with \textit{CXO}, provided that its proper motion is smaller than $\sim$0.08/10=0.008\arcsec\,yr$^{-1}$$\cong$2350\,km\,s$^{-1}$. 

\begin{figure*}[htb!]
\centerline{\includegraphics[scale=0.5]{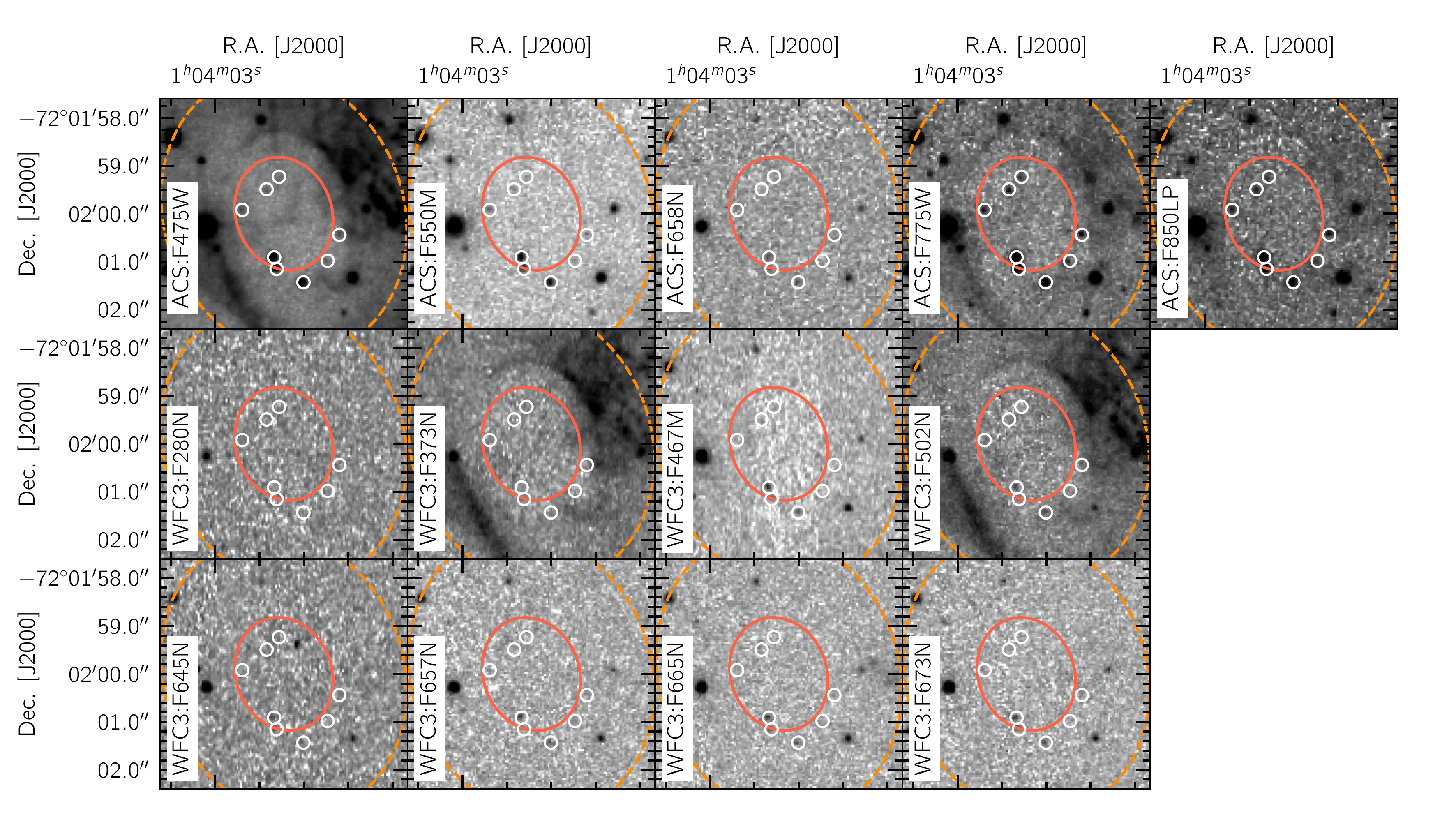}}
\caption{\linespread{1.0}\selectfont{}Detailed views of the torus region in archival \textit{HST} ACS and WFC3 observations. The spatial extent of the optical ring detected with MUSE is traced using the ellipses introduced in the main article. Point-sources detected near/within the inner ellipse (in at least one image) are marked with white circles. We find no candidate near the center of the ring. The presence of the filamentary structure of ejecta is visible in the F475W (via \oiii\,$\lambda\lambda$4959,5007), F775W (via \oi\,$\lambda$7774), F850LP (via \oi\,$\lambda\lambda$9262,9266), F373N (via \oii\,$\lambda\lambda$3726,3728) and F502N (via \oiii\,$\lambda\lambda$4959,5007) images.}\label{fig:hst}
\end{figure*}

Postage stamp images of the optical ring area detected with MUSE, for all re-processed \textit{HST} camera+filter combinations, are presented in Fig.~\ref{fig:hst}. We find no evidence for an optical counterpart to the X-ray point source detected with \textit{CXO}. Four point sources are detected within the central gap of the low-ionization optical ring detected with MUSE, but we rule them out as suitable candidate based on their largely off-center locations. \1E0102 was not observed by \textit{HST} down to the same depth in each filter. The strongest constraints for the brightness of a possible optical counterpart come from the ACS F475W image (4324\,s on-source), the ACS F775W image (2160\,s on-source), and the ACS F850LP (2160\,s on-source). Specifically, we derive an upper limit for the magnitude of a possible optical counterpart to the X-ray source of m$_\textrm{F475W}>23.0$, m$_\textrm{F775W}>24.5$ and m$_\textrm{F850LP}>25.0$, noting that the F475W image is severely affected by contaminating \oiii\,$\lambda\lambda$4959,5007 emission from the fast ejecta, falling within the filter bandpass.

\subsection*{\textit{CXO} characterization of the CCO in \1E0102}
\subsubsection*{Spatial characterization}\label{sec:cxo-spatial}

Combined X-ray images of \1E0102 were created in the standard \textit{CXO} ACIS source detection energy bands, as well as the ultrasoft band (0.2--0.5 keV), using the \textsc{ciao} script \textsc{merge\_obs}. We ran the Mexican-Hat wavelet source detection tool \textsc{wavdetect} on all the combined images, which confirmed the presence of the point source \textit{p1} already detected by \cite{Rutkowski2010}, and coincident with the low-ionization optical ring discovered with MUSE. As we are exclusively searching for a point source embedded in the diffuse emission of the hot ejecta from the SNR, the source detection algorithm was run with no point spread function (PSF) map on ``fine'' scales (the \textsc{wavdetect} scales parameter was set to ``1 2''). The derived spatial parameters for each band are summarized in Table~\ref{table:cxo_image}.

\begin{table*}
\begin{center}
{
\caption{ \linespread{1.0}\selectfont{} Spatial characterization of the point source \textit{p1} detected in \1E0102 by \textit{CXO}, derived from \textsc{wavdetect}.}\label{table:cxo_image}
\smaller
\begin{tabular}{lccccccc}
\hline
Band & Energy & R.A. & Dec. & Semi-major axis & Semi-minor axis & Pos. Angle	& Detection significance \\
	 & [keV]  & [J2000] & [J2000] & [arcsec] & [arcsec]	& [deg] & \\
\hline
Broad 	    & 0.5--7.0 & 01:04:02.73 & -72:02:00.36 & 1.6 & 1.1 & 50.9	& 8.60 \\
Ultrasoft	& 0.2--0.5 & 01:04:02.73 & -72:02:00.23	& 1.2 & 1.0	& 58.5 & 2.10 \\
Soft		& 0.5--1.2 & 01:04:02.77 & -72:02:00.66	& 1.9 & 0.9	& 74.2 & 5.78 \\
Medium	    & 1.2--2.0 & 01:04:02.75 & -72:02:00.14	& 1.2 & 1.0	& 54.0 & 8.30 \\
Hard		& 2.0--7.0 & 01:04:02.75 & -72:02:00.19	& 2.2 & 1.8	& 91.8 & 5.45 \\
\hline
\end{tabular}}
\end{center}
\end{table*}

Without a PSF map, \textsc{wavdetect} uses the smallest wavelet scale found to derive the source properties. This approach may lead to incorrect parameters at large off-axis angles, but our \textit{target on-axis} selection criteria for the \textit{CXO} datasets ensures that this will not have a very significant effect in the present case. The detection process itself is unaffected by the lack of inclusion of a PSF map. We did experiment with PSF maps weighted by both exposure time and exposure map for the combined broad band dataset. But in each instance, the algorithm detects the entire filamentary structure of \1E0102, rather than a point source itself\footnote{For more details on running \textsc{wavedetect} on combined datasets, see \textit{http://cxc.harvard.edu/ciao/threads/wavdetect merged/index.html}}.

\subsubsection*{Spectral characterization}

We manually extracted the counts of the source \textit{p1} in all bands using the region derived from the automated point source search in the medium \textit{CXO} band, where it is most reliably detected by \textsc{wavdetect} (see Table~\ref{table:cxo_image}). Background counts are derived from an annulus centered on the source, and with inner and outer radii of 1.2\arcsec\ and 1.8\arcsec, respectively. The counts derived from each band are summarized in Table~\ref{tab:counts}.

\begin{table*}
\centering
\caption{\linespread{1.0}\selectfont{} X-ray properties of the point source \textit{p1}, derived from our best-fit, absorbed, single blackbody model with $kT_{BB}=0.19$~keV. The error for the number counts and corresponding fluxes correspond to the 1-$\sigma$ Poisson errors on the X-ray number counts of the source \textit{p1} in the different energy ranges.The errors associated with luminosity $L_i$ stem from an assumed 10\% uncertainty in the depth of \1E0102 within the SMC.\newline}\label{tab:counts} 
\begin{threeparttable}
\centering
{\smaller
\renewcommand{\arraystretch}{0.7}
\begin{tabular}{lcccccc}
\hline\\\noalign{\smallskip}
Band & Energy & Counts$_\textrm{observed}$\tnote{(a)} & Counts$_\textrm{simulated}$\tnote{(b)} & $F_{-14,\textrm{ model}}$\tnote{(c)} & $F_{-14,\textrm{ data}}$\tnote{(d)} & $L_{i,33,\textrm{model}}$\tnote{(e)} \\\noalign{\smallskip}
	& [keV]	& &	& [erg~cm$^{-2}$~s$^{-1}$] &  [erg~cm$^{-2}$~s$^{-1}$]	& [erg~s$^{-1}$] \\\noalign{\smallskip}
\hline\\\noalign{\smallskip}
Broad & 0.5--7.0 & $469\pm56$ & 469 & 1.37 &$1.4\pm0.2$	& $9.0\pm1.3$\\\noalign{\smallskip}
Ultrasoft & 0.2--0.5	& $<40$\tnote{(f)} & 4 & 0.17 & $<2.1$\tnote{(f)} & $3.2\pm0.4$ \\\noalign{\smallskip}
Soft & 0.5--1.2	& $287\pm51$ & 279 & 1.11 & $1.1\pm0.2$ & $7.5\pm1.1$ \\\noalign{\smallskip}	
Medium & 1.2--2.0 & $143\pm22$ & 172 & 0.28 & $0.24\pm0.04$ & $1.4\pm0.2$ \\\noalign{\smallskip}
Hard & 2.0--7.0	& $39\pm9$ & 7 & 0.02 & $0.10\pm0.02$ & $0.08\pm0.01$ \\\noalign{\smallskip}
\hline\\\noalign{\smallskip}
\end{tabular}}
\begin{tablenotes}
\footnotesize
\item [(a)] Observed (background subtracted) counts for the source \textit{p1}.
\item [(b)] Simulated source counts for the best-fit, absorbed single blackbody model ($wabs\times wabs \times bbody$, with $kT_{BB}=0.19$~keV).
\item [(c)] The X-ray flux (in units of $10^{-14}$ erg~cm$^{-2}$~s$^{-1}$) of the best-fit, absorbed single blackbody model.
\item [(d)] The observed X-ray flux (in units of $10^{-14}$ erg~cm$^{-2}$~s$^{-1}$) of the source \textit{p1}, computed with the ECFs derived (for each energy band) from the best-fit, absorbed single blackbody model and associated simulated dataset.
\item [(e)] Intrinsic (i.e. unabsorbed) X-ray luminosity implied from the assumed absorbed Blackbody model, assuming a distance of 62.1~kpc to the SMC and a 10\% error on the exact distance to \1E0102 within the SMC.
\item [(f)] 3$\sigma$ upper limit.
\end{tablenotes}
\end{threeparttable}
\end{table*}

Our chosen parameters for the ``background annulus'' are motivated by the complex X-ray background emission throughout \1E0102. This annulus is narrow enough to avoid brighter filamentary structures nearby, but large enough for a reliable estimate of the local background level close to and around the source. Whilst our region size is small, approximately 90\% of the encircled energy still lies within 1\arcsec\ of the central pixel\footnote{See Figure 6.10 on \\ \textit{http://cxc.cfa.harvard.edu/proposer/POG/html/ACIS.html}}. \cite{Pavlov2009} also adopt a similar region size (1.5\arcsec\ in radius) in their analysis of the Galactic CCO in Cas A. We note that we find no evidence for the background annuli to contain any knots or feature associated with the Ne and Mg metal lines, after the visual inspection of dedicated channel maps centered at 0.8496\,keV and 1.2536\,keV, each 130\,eV wide.
 
The low counts associated with source \textit{p1} hinder our ability to extract robust parameters from a direct spectral fitting. Instead, for comparison purposes, we first investigate how the CCO of Cas A would appear to an observer if it were located at the distance of the SMC, with an absorbing column equal to that of \1E0102. We simulate the double-blackbody model of \cite{Pavlov2009} to do so, using \textsc{xspec} v.12.9.0 and the \textsc{fakeit} command. These authors fit absorbed blackbody, power law and neutron star atmosphere models to a single 70.2 ks \textit{CXO} observation of Cas A, with a count rate of 0.1 counts\,s$^{-1}$. They find that the X-ray spectrum of the CCO of Cas A is equally well described by both an absorbed double neutron star model (represented as $wabs\times(nsa+nsa)$ in \textsc{xspec}) and an absorbed double blackbody model (represented as $wabs\times(bbody+bbody)$ in \textsc{xspec}). We voluntarily reproduce only the double blackbody model here, as the double neutron star model has many more parameters which are not explicitly described.

We include two absorption terms to the model: one to account for the Galactic absorption \citep[set to $5.36\times10^{20}\textrm{~cm}^{-2}$;][]{Dickey1990}, and another set to the intrinsic absorption in the south-east region of \1E0102, where the source \textit{p1} is located \citep[$4\times10^{20}\textrm{~cm}^{-2}$;][]{Sasaki2006}. The spectrum is then normalised to have the same unabsorbed luminosity as that reported by \cite{Pavlov2009}, over the same energy range. In summary, our final model is represented by $C\times wabs_{Gal}\times wabs_i\times(bbody+bbody)$ in \textsc{xspec}, where $C$ is a constant. We simulate the spectrum of such a source using the \emph{CXO} cycle 19 canned response matrices\footnote{Available at \textit{http://cxc.harvard.edu/caldb/prop\_plan/imaging/index.html}} and an exposure time of 323\,ks, matching the total depth of our combined \textit{CXO} observations. 

We compare the simulated spectrum of Cas A with the measured counts of the source \textit{p1} (in each of the narrow energy bands) in the top panel of Fig.~\ref{fig:xray}: \textit{p1} appears cooler than would a Cas A-like CCO at the same location. This is not necessarily surprising, given that \1E0102 is $\sim$1700\,yr older than Cas A.

\begin{figure*}
\centerline{\includegraphics[angle=90,width=0.75\textwidth]{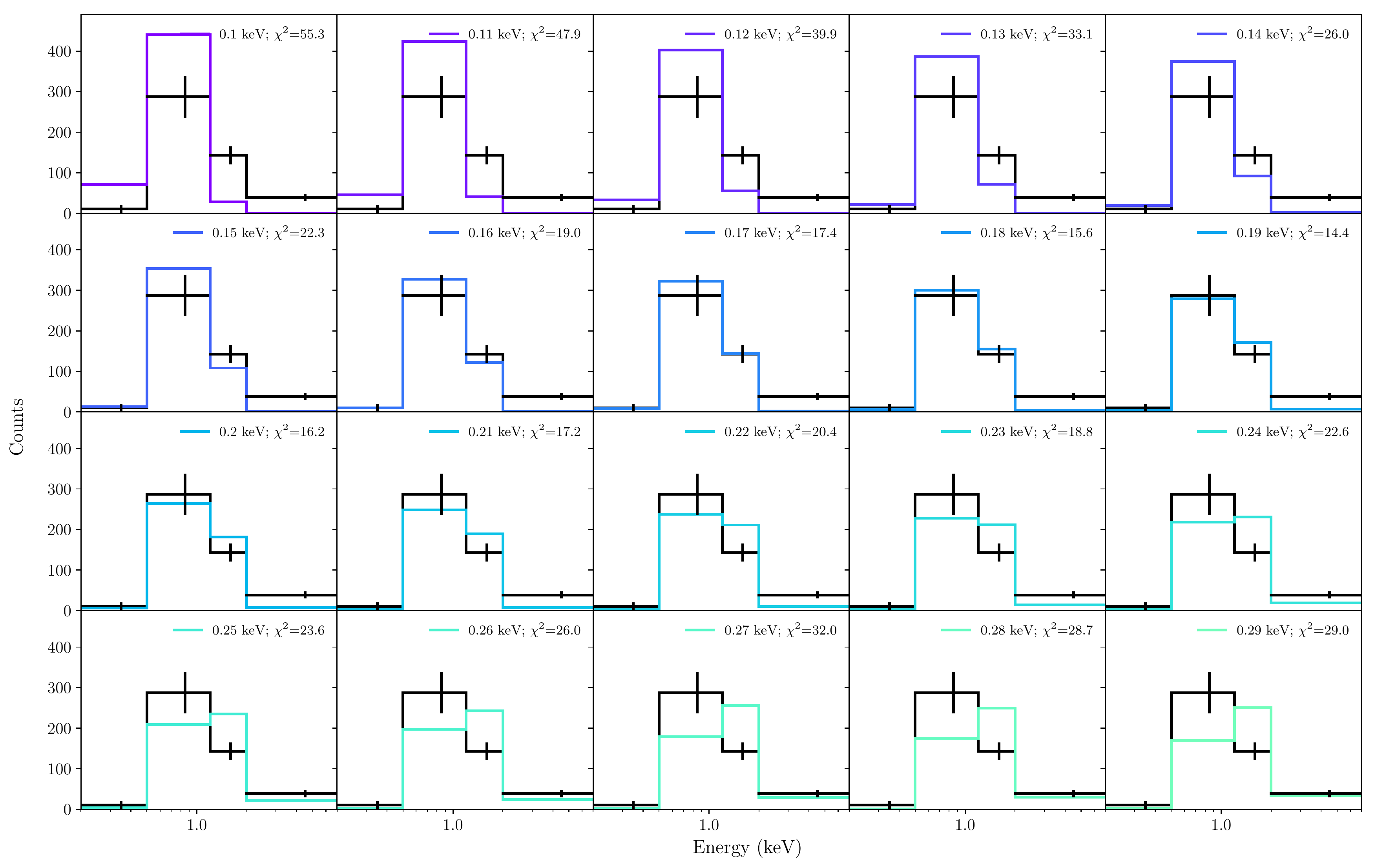}}
\caption{\linespread{1.0}\selectfont{} Energy distribution of the X-ray point source \textit{p1} (black curves), compared with absorbed, single blackbody models (colored curves) with $kT_\mathrm{BB}$ ranging from 0.10--0.29~keV in steps of 0.01~keV. The $\chi^2$ for each model, computed over the 0.5--0.7\,keV energy range, is shown in each panel. The error bars correspond to the 1-$\sigma$ Poisson errors on the X-ray number counts of the source \textit{p1}.}\label{fig:xray-models1}
\end{figure*}

\begin{figure*}
\centerline{\includegraphics[angle=90,width=0.75\textwidth]{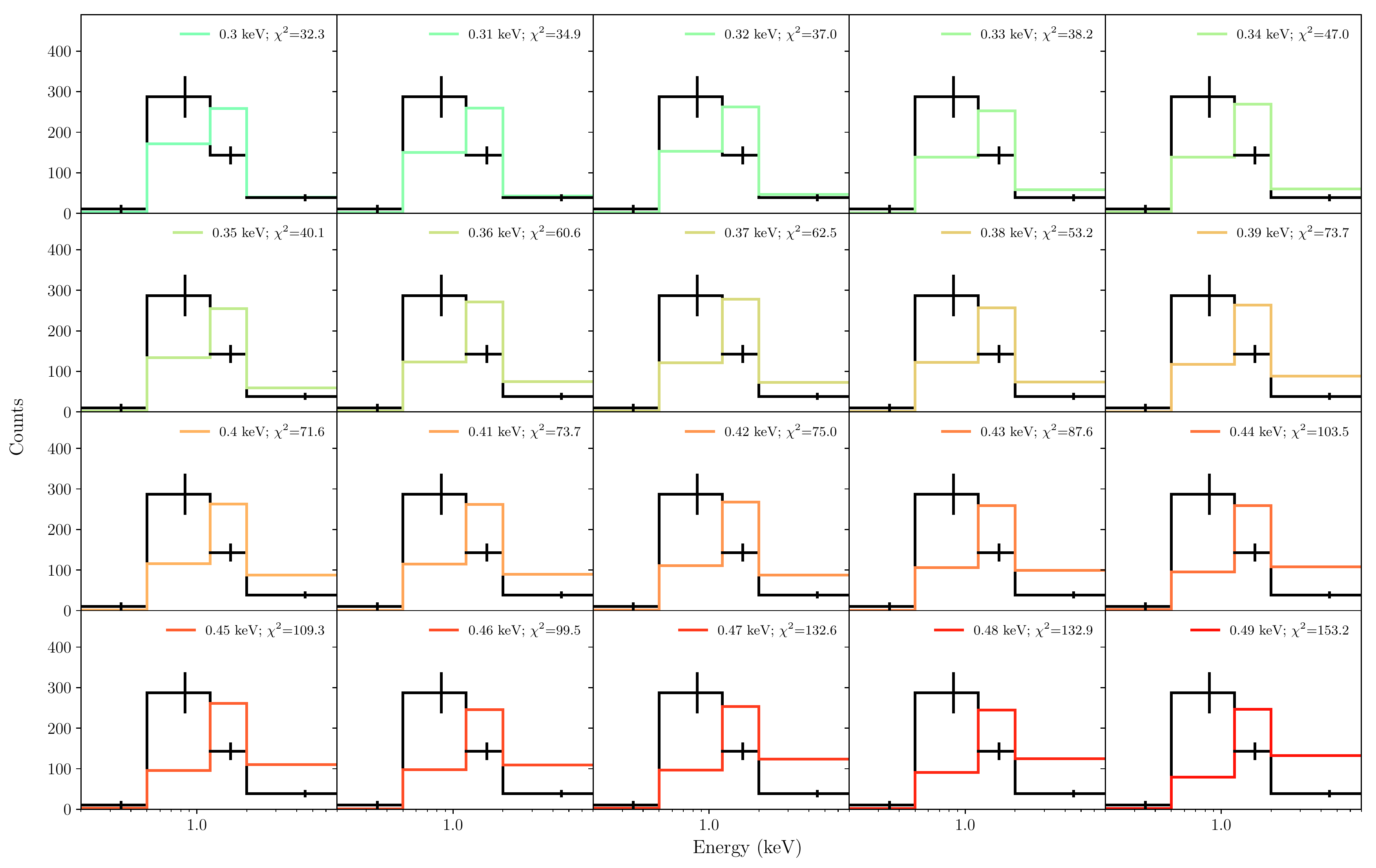}}
\caption{Same as Fig.~\ref{fig:xray-models1}, but for models with $kT_\mathrm{BB}$ ranging from 0.30--0.49~keV.}\label{fig:xray-models2}
\end{figure*}

\begin{figure*}
\vspace{-75pt}
\centerline{\includegraphics[angle=90,width=0.75\textwidth]{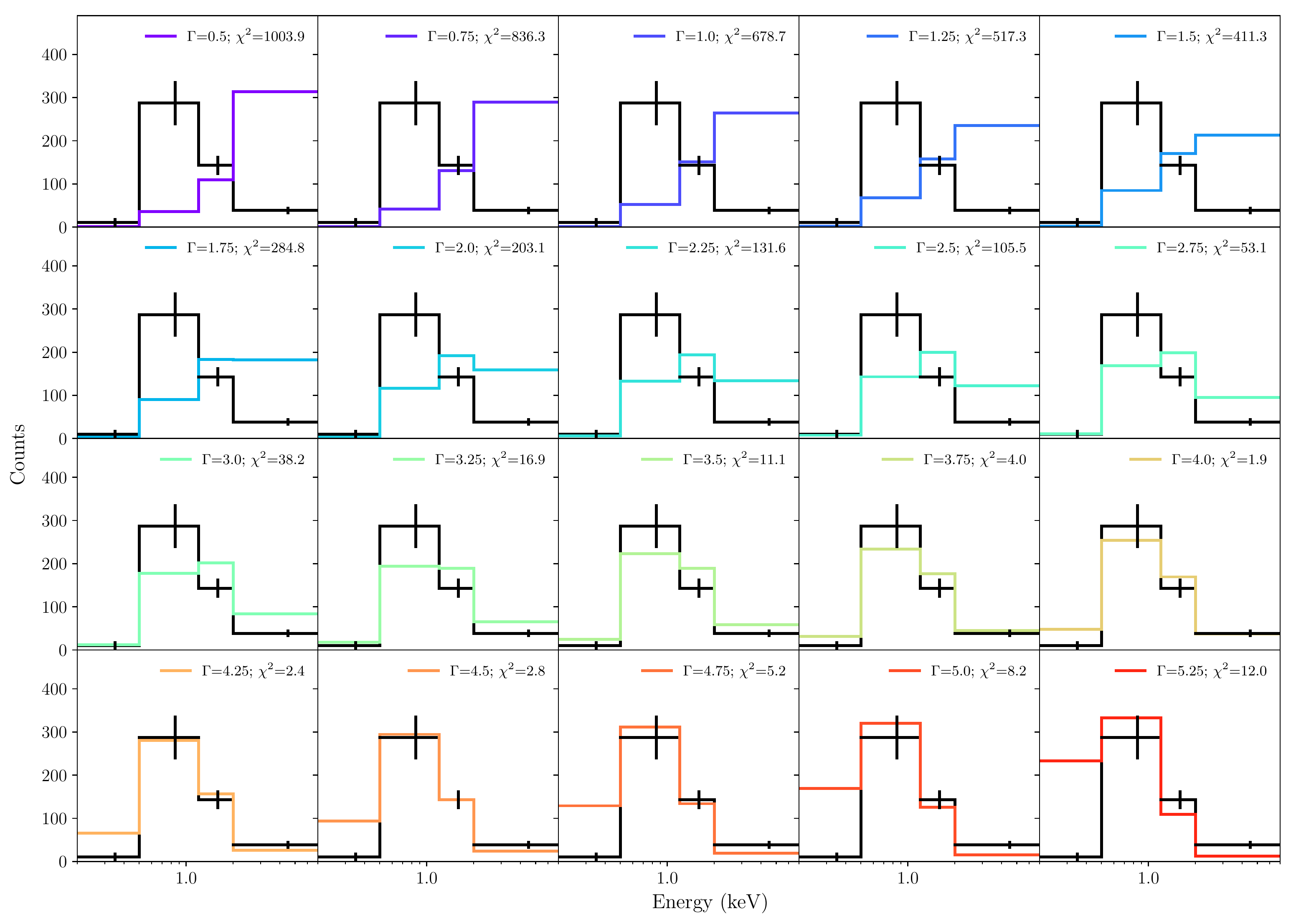}}
\caption{ \linespread{1.0}\selectfont{} Energy distribution of the X-ray point source \textit{p1} (black curves), compared with absorbed, power law models (colored curves) with photon index $\Gamma$ ranging from 0.50--5.25 in steps of 0.25. The $\chi^2$ for each model, computed over the 0.5--0.7\,keV energy range, is shown in each panel. The error bars correspond to the 1-$\sigma$ Poisson errors on the X-ray number counts of the source \textit{p1}.}\label{fig:xray-models3}
\end{figure*}

Next, to derive X-ray fluxes and luminosities for the source \textit{p1} in \1E0102, we use \textsc{xspec}'s \textsc{fakeit} with the cycle 19 canned response matrices to simulate a) absorbed single blackbody models and b) absorbed power laws in the same manner as above. These models are represented by $C\times wabs_\textrm{Gal}\times wabs_i\times bbody$ and $C\times wabs_\textrm{Gal}\times wabs_i\times powerlaw$ in \textsc{xspec}. We restrict ourselves to single blackbody and power law models, as the counts associated with the source \textit{p1} are insufficient to reliably constraint more complex ones, including double blackbody models. The Galactic and intrinsic absorption components are once again set to $5.36\times10^{20}\textrm{~cm}^{-2}$ and $4\times10^{20}\textrm{~cm}^{-2}$, respectively. We generate spectra with temperatures $kT_\mathrm{BB}$ ranging from 0.10--0.49~keV in steps of 0.01~keV for the blackbody models, and with a photon index $\Gamma$ ranging from 0.50--5.25 in steps of 0.25 for the power law models. Each spectrum is normalised so that the total number of counts over 0.5--7.0~keV (i.e. the broad band) in a 323~ks exposure is consistent with that of the source \textit{p1} detected in our merged data set of \1E0102. All our models are compared with the measured counts of the source \textit{p1} in Fig.~\ref{fig:xray-models1} to \ref{fig:xray-models3}.

\begin{figure}
\centerline{\includegraphics[width=0.45\textwidth]{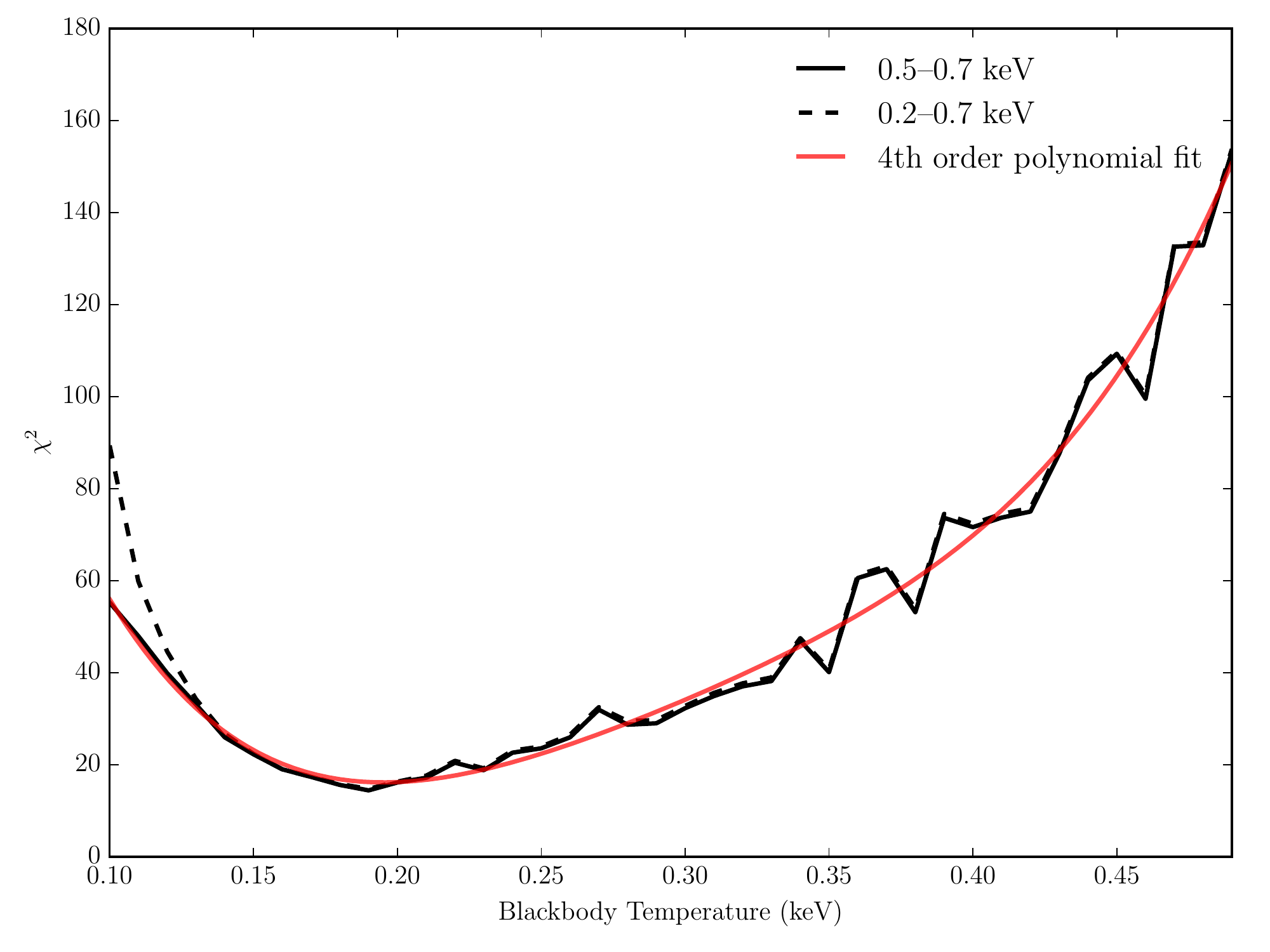}}
\caption{ \linespread{1.0}\selectfont{} The $\chi^2$ values for each single blackbody model fitted to the energy distribution of the point source \textit{p1}, shown as a function of the model temperature and computed over the 0.5--0.7\,keV (full black line) and 0.2--0.7\,keV (dashed black line) energy range. The best-fit fourth-order polynomial to the 0.5--0.7\,keV dataset is shown with a red line.}\label{fig:chi}
\end{figure}

We calculate the $\chi^2$ of each of the single blackbody models with respect to the data over both the 0.5--7.0~keV and 0.2--7.0~keV energy ranges, and for each of the power law models with respect to the data over the 0.5--7.0~keV energy range. We exclude the ultrasoft band from some of these ranges, as the \textit{CXO} response files are less accurate below 0.5~keV\footnote{see \textit{http://cxc.harvard.edu/caldb/prop\_plan/imaging/index.html}}. We show in Fig.~\ref{fig:chi} the distribution of these $\chi^2$ values as a function of the blackbody temperature of the models, together with the fourth order polynomial best fit to the 0.5--0.7~keV data. For the absorbed, single blackbody models, we find that the source \textit{p1} is best characterised by a blackbody with $kT_\mathrm{BB}~0.19\pm0.02$~keV. For the absorbed, power law models, we find that the source \textit{p1} is best characterised by models with a photon index $3.75<\Gamma<4.75$.

Fitting a single blackbody model to Cas A (with a derived a temperature of 0.4~keV) underestimates the CCO flux at higher energies \citep{Pavlov2009}. The same can be seen in our simulations: our best-match, absorbed single blackbody model underestimate the number counts of the source \textit{p1} in the hard \textit{CXO} band. The flux values associated with the best-fit $kT_\mathrm{BB}=0.19$~keV model are included in Table~\ref{tab:counts} along with the implied intrinsic (i.e. un-absorbed) X-ray luminosity. The errors associated with the latter stem from an assumed 10\% uncertainty in the depth of \1E0102 within the SMC. In the same Table, we also present the observed X-ray flux of the source \textit{p1}, computed from the observed data using the energy conversion factors (ECFs) derived (for each energy band) from the best-fit, absorbed single blackbody model with $kT_\mathrm{BB}=0.19$~keV.

We note that our derived fluxes and luminosities are subject to caveats. First, our simulated spectra assume that we have a single 323\,ks observation performed in cycle 19 of \textit{CXO}'s lifetime, rather than several observations spanning over a decade. Second, the background subtraction and determination for the point source \textit{p1} is non-trival, as it is embedded in bright, spatially complex background emission. A more complex spectroscopic analysis of this X-ray point source, outside the scope of this report, ought to address these points carefully.

\subsection*{MUSE spectral characterization of the optical ring}\label{sec:specfit}

\begin{figure*}[htb!]
\vspace{0pt}
\centerline{\includegraphics[width=\textwidth]{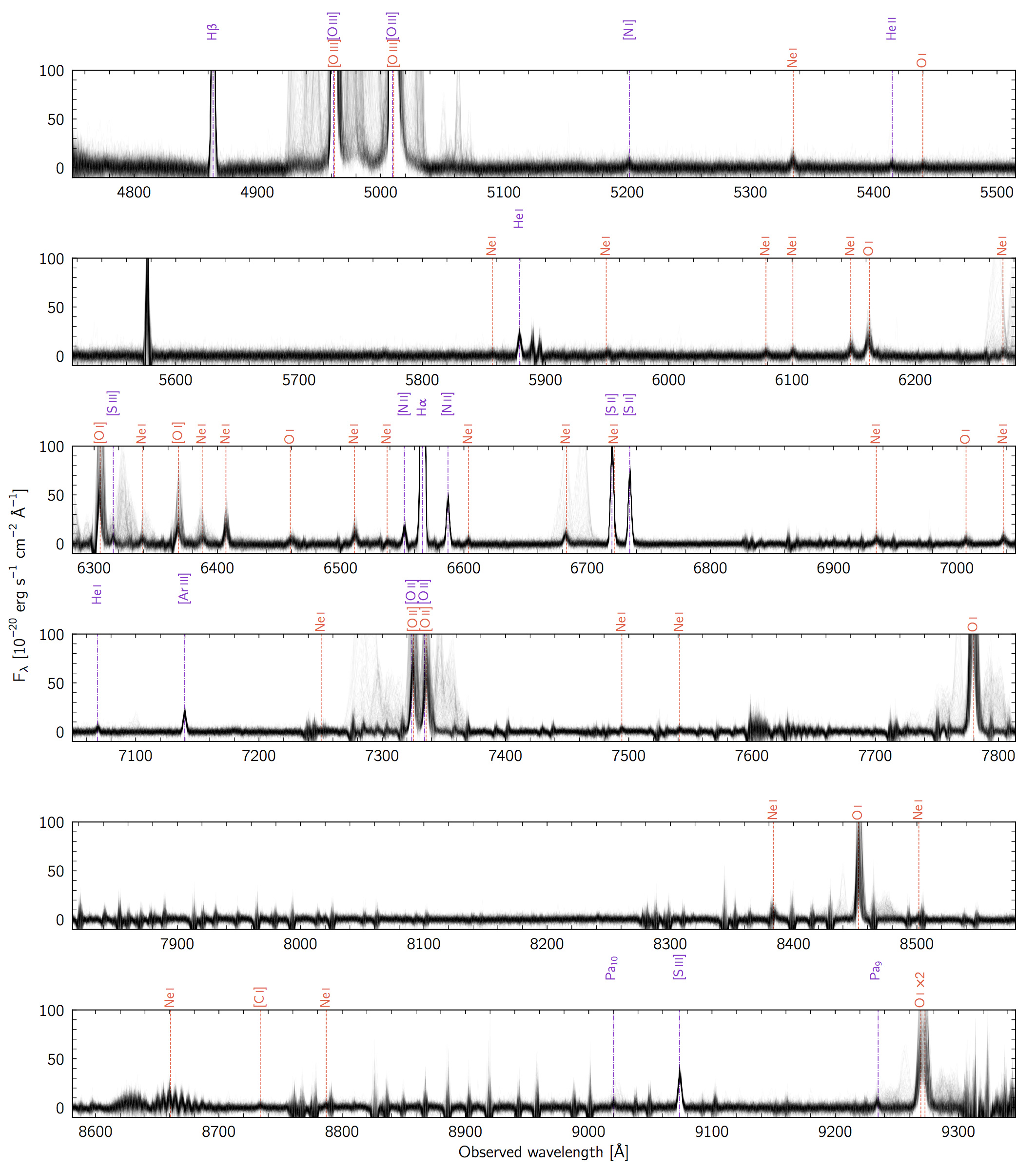}}
\caption{\linespread{1.0}\selectfont{}MUSE spectra for all 492 spaxels within the footprint of the low-ionization ring, shown as individual black lines with 5\% transparency. Light-grey spectral regions indicate spatially-varying emission lines, associated with reverse-shocked fast ejecta (the most prominent example of which is associated with \oiii\,$\lambda\lambda$4959,5007). Darker, sharper emission lines are associated with the torus (and labelled using red dashed lines) or with background H\,\textsc{\smaller II}-like emission from the SMC (the most prominent examples of which are labelled using dot-dashed purple lines). Any other spectral feature is caused by the imperfect sky subtraction of the data (either for sky emission lines or telluric lines).} \label{fig:spec}
\end{figure*}

We performed a full fit of the MUSE spectra associated with each of the 492 spaxels contained within the area of the optical ring. The fitted spaxels are located within two ellipses with semi-major axis of 1.2\arcsec\ and 3.0\arcsec\, centered on the CCO (R.A: 01$^h$04$^m$02.7$^s$ | Dec: -72$^{\circ}$02$^{\prime}$00.2$^{\prime\prime}$ [J2000]), and rotated by 20$^{\circ}$ East-of-North. The ring is detected in 35 lines of \nei, \oi, [\oi], and \ci, all of which were identified following a manual, visual inspection of every spectral channel in the continuum-subtracted MUSE datacube (see Fig.~\ref{fig:spec}). An intensity increase spatially coincident with the ring is also detected in 4 lines of \oiii\ and \oii, but the surrounding emission of these lines is spatially more extended in comparison to the lower-ionization lines, as illustrated in Fig.~\ref{fig:zoom}. The low- and high-ionization lines thus likely originate from physically distinct volumes.

The line fitting is performed using a custom routine relying on the \textsc{python} implementation of \textsc{mpfit}, a script that uses the Levenberg-Marquardt technique \citep{More1978} to solve least-squares problems, based on an original Fortran code part of the \textsc{minpack}-1 package. We do not include the \oiii\,$\lambda\lambda$4959,5007 lines in the fitting: their large intensity with respect to the lower-ionization emission lines and the presence of several fast ejecta knots throughout the footprint of the ring affect the derived kinematic signature of the ring. We do however include the \oii\,$\lambda\lambda$7320,7330 lines in the list of fitted lines for comparison purposes, after having ensured that their inclusion does not affect the outcome of the spectral fit at a significant level. Finally, we do not include the \nei\,$\lambda$6266, \nei\,$\lambda$6334, \nei\,$\lambda$6383, \nei\,$\lambda$6678, \nei\,$\lambda$6717, and \nei\,$\lambda$8654 emission lines in the fit, as these lines are strongly contaminated by residual sky lines, bright emission lines from fast ejecta along specific lines-of-sight, and/or background H\,\textsc{\smaller II}-like emission.

For each of the 492 spaxels within the footprint of the ring, the selected 31 emission lines are fitted simultaneously, each with a single Gaussian component tied to a common, observed LOS velocity $v_\textrm{los,obs}$. The velocity dispersion $\sigma(\lambda)$ for a specific line wavelength $\lambda$ is set to:
\begin{equation}
\sigma(\lambda)= \sqrt{\sigma_\textrm{obs}^2+\sigma_\textrm{inst}(\lambda)^2},
\end{equation}
with $\sigma_\textrm{obs}=65$ km\,s$^{-1}$ the assumed-constant underlying gas velocity dispersion, and $\sigma_\textrm{inst}(\lambda)$ the wavelength-dependant instrumental spectral dispersion of MUSE. The ring emission lines are spectrally unresolved by MUSE. Here we fix their underlying velocity dispersion to ensure a robust fit for all spaxels, including those towards the inner and outer edges of the ring with lower S/N, noting that the fitting results are not significantly affected by the exact value of $\sigma_\textrm{obs}$. Dedicated observations with higher spectral resolution are required to constrain this parameter. 

The measured line fluxes for all fitted lines are presented in Table~\ref{table:spec}. The rest frame line wavelengths were obtained from the NIST Atomic Spectra Database \citep{Kramida2016}. The LOS velocity map of the ring is shown in Fig.~\ref{fig:torus_RGB}, together with a pseudo-RGB image of the fitted area in the light of \nei\,$\lambda$6402, \oii\,$\lambda$7330, and \oi\,$\lambda$7774. The mean fitted ring spectrum is compared to the observed spectra in Fig.~\ref{fig:spec}. The mean velocity of the optical ring is $(55\pm10)$ km\,s$^{-1}$ with respect to the background H\,\textsc{\smaller II}-like emission of the SMC. The uncertainty is dominated by the somewhat irregular appearance of the velocity map: the error associated with the fitting alone is of the order of 5\,km\,s$^{-1}$. 

\begin{table*}
\begin{center}
\vspace{-75pt}
{
\caption{\linespread{1.0}\selectfont{}Total line fluxes $F_{\lambda,tot}$, average line flux densities $<F_\lambda>$, and associated 1-$\sigma$ standard deviation $\sigma(F_\lambda)$ of the optical ring surrounding the CCO in \1E0102. The quoted errors for the line flux and line flux density are at the 1-$\sigma$ level. The exhaustive list of ring emission lines were identified by a visual, manual inspection of each spectral channel of the MUSE datacube. The line fluxes are derived from a simultaneous spectral fit on a spaxel-by-spaxel basis, for each of the 492 MUSE spaxels within the 19.7 square arcseconds footprint of the ring.}\label{table:spec}
\smaller
\begin{tabular}{c c c c c}
\hline
Line & $\lambda_\textrm{rest}$ & $F_{\lambda,tot}$ & $<F_\lambda>$ & $\sigma(F_\lambda)$ \\
     & [\AA] & [$10^{-18}$ erg s$^{-1}$ cm$^{-2}$] & [$10^{-20}$ erg s$^{-1}$ cm$^{-2}$ arcsec$^{-2}$] & [$10^{-20}$ erg s$^{-1}$ cm$^{-2}$ arcsec$^{-2}$] \\
\hline\hline
\oiii & 4958.911 & N/A & N/A & N/A \\
\oiii & 5006.843 & N/A & N/A & N/A \\
\nei & 5330.7775 & $136.6\pm2.3$ & $694.2\pm11.9$ & 460.4 \\
\oi  & 5435.78   & $39.7\pm1.8$ & $202.0\pm9.0$ & 236.4\\
\nei & 5852.4878 & $38.1\pm1.7$ & $193.5\pm8.5$ & 213.6\\
\nei & 5944.8340 & $49.1\pm1.9$ & $249.7\pm9.5$ & 249.3\\
\nei & 6074.3376 & $60.5\pm1.7$ & $307.3\pm8.6$ & 263.6\\
\nei & 6096.1630 & $70.7\pm1.8$ & $359.4\pm9.1$ & 286.9\\
\nei & 6143.0627 & $166.2\pm1.9$ & $844.7\pm9.8$ & 546.8\\
\oi  & 6158.18   & $398.5\pm2.0$ & $2024.7\pm10.3$ & 886.9\\
\nei & 6266.4952 & N/A & N/A & N/A \\
$[$\oi$]$& 6300.304 & $1799.1\pm3.1$ & $9142.0\pm15.7$ & 5858.3\\
\nei & 6334.4276 & N/A & N/A & N/A \\
$[$\oi$]$& 6363.776 & $566.2\pm2.1$ & $2877.3\pm10.6$ & 1943.9\\
\nei & 6382.9914 & N/A & N/A & N/A \\
\nei & 6402.248  & $373.6\pm1.9$ & $1875.3\pm9.4$ & 736.4\\
\oi  & 6454.44   & $78.6\pm1.7$ & $399.2\pm8.5$ & 303.4\\
\nei & 6506.5277 & $219.2\pm1.7$ & $1113.6\pm8.9$ & 561.7\\
\nei & 6532.8824 & $29.0\pm1.3$ & $147.6\pm6.5$ & 170.8\\
\nei & 6598.9528 & $34.3\pm1.5$ & $174.4\pm7.7$ & 184.2\\
\nei & 6678.2766 & N/A & N/A & N/A \\
\nei & 6717.0430 & N/A & N/A & N/A  \\
\nei & 6929.4672 & $101.6\pm1.4$ & $516.3\pm7.1$ & 289.4\\
\oi  & 7002.23   & $73.6\pm1.3$ & $374.2\pm6.8$ & 301.6\\
\nei & 7032.4128 & $113.5\pm1.4$ & $576.9\pm6.9$ & 303.8\\
\nei & 7245.1665 & $30.3\pm1.5$ & $153.7\pm7.4$ & 189.7\\
\oii & 7319.92   & $2357.5\pm2.4$ & $11979.1\pm12.1$ & 7512.0\\
\oii & 7330.19   & $1904.3\pm2.2$ & $9676.5\pm11.1$ & 5273.2\\
\nei & 7488.8712 & $43.3\pm1.4$ & $220.1\pm7.0$ & 181.5\\
\nei & 7535.7739 & $41.1\pm1.3$ & $208.8\pm6.8$ & 179.8\\
\oi  & 7774.17   & $4442.1\pm3.5$ & $22571.6\pm17.6$ & 9614.1\\
\nei & 8377.6070 & $147.0\pm2.3$ & $747.0\pm11.7$ & 469.5\\
\oi  & 8446.36   & $1419.1\pm3.3$ & $7210.8\pm16.7$ & 3949.6\\
\nei & 8495.3591 & $36.7\pm1.7$ & $186.6\pm8.5$ & 227.9\\
\nei & 8654.3828 & N/A & N/A & N/A \\
\ci  & 8727.13   & $44.9\pm1.2$ & $228.4\pm6.3$ & 211.1\\
\nei & 8780.6223 & $49.8\pm1.6$ & $252.9\pm8.2$ & 259.2\\
\oi  & 9262.67   & $1246.7\pm3.7$ & $6334.7\pm18.7$ & 2758.9\\
\oi  & 9266.01   & $1338.7\pm3.7$ & $6802.4\pm18.9$ & 3285.6\\
\hline
\end{tabular}}
\end{center}
\end{table*}

\section*{Data Availability Statement} The data that support the plots within this paper and other findings of this study are available from the corresponding author upon reasonable request. The MUSE observations are available from the ESO archive under program ID 297.D-5058. The \textit{HST} observations are available from the Barbara A. Mikulski Archive for the Space Telescopes and the \textit{CXO} observations from the Chandra Data Archive.

\end{multicols}

\end{document}